\documentclass[%
 reprint,
superscriptaddress,
nofootinbib,
 amsmath,amssymb,
aps, prl
longbibliography
]{revtex4-2}
\usepackage{mathtools}
\usepackage{makecell} %
\usepackage{subfiles}
\usepackage{graphicx}
\usepackage{bm}
\usepackage{color}
\usepackage[ruled, linesnumbered, boxruled]{algorithm2e}
\usepackage{booktabs}
\usepackage{breakurl}

\begin{document}
\title{Towards the ground state of molecules via diffusion Monte Carlo on neural networks}

\author{Weiluo Ren}
\email{renweiluo@bytedance.com}
\affiliation{ByteDance Research, Zhonghang Plaza, No. 43,  North 3rd Ring West Road, Haidian District, Beijing, People’s Republic of China}

\author{Weizhong Fu}
\thanks{W.R. and W.F. contributed equally to this work.}
\affiliation{ByteDance Research, Zhonghang Plaza, No. 43,  North 3rd Ring West Road, Haidian District, Beijing, People’s Republic of China}
\affiliation{School of Physics, Peking University, Beijing 100871, People’s Republic of China}
\author{Xiaojie Wu}
\affiliation{ByteDance Research, Zhonghang Plaza, No. 43,  North 3rd Ring West Road, Haidian District, Beijing, People’s Republic of China}
\author{Ji Chen}
\email{ji.chen@pku.edu.cn}
\affiliation{School of Physics, Peking University, Beijing 100871, People’s Republic of China}
\affiliation{Interdisciplinary Institute of Light-Element Quantum Materials, Frontiers
Science Center for Nano-Optoelectronics, Peking University, Beijing 100871, People’s Republic of China}

\date{\today}

\begin{abstract}
Diffusion Monte Carlo (DMC) based on fixed-node approximation has enjoyed significant developments in the past decades and become one of the go-to methods when accurate ground state energy of molecules and materials is needed. 
However, the inaccurate nodal structure hinders the application of DMC for more challenging electronic correlation problems.
In this work, we apply the neural-network based trial wavefunction in fixed-node DMC, which allows accurate calculations of a broad range of atomic and molecular systems of different electronic characteristics.
Our method is superior in both accuracy and efficiency compared to state-of-the-art neural network methods using variational Monte Carlo (VMC).
We also introduce an extrapolation scheme based on the empirical linearity between VMC and DMC energies, and significantly improve our binding energy calculation.
Overall, this computational framework provides a benchmark for accurate solutions of correlated electronic wavefunction and also sheds light on the chemical understanding of molecules.
\end{abstract}

\maketitle

\section{Introduction}
\label{sec:intro}

Since the establishment of quantum wavefunction theory by Erwin Schr{\"o}dinger, \textit{ab initio} electronic structure calculation has become  one of the holy grails in chemistry \cite{PopleNobel,KohnNobel}. 
Molecules generally consist of a set of nuclei bonded together via electrons through electrostatic interactions. 
Therefore, the ground state electronic structure, i.e. the many body electronic wavefunction, is very much the most fundamental property, based on which we form the basic understanding of molecules.
On top of the ground state wavefunction solution, one may further study electronic excitation, calculate nuclear forces and vibrations, optimize molecular structures, model dynamics and reactions, etc. \cite{Helgaker_recent_2012}.
Approximated methods, such as density functional theory and post Hartree-Fock methods have been widely employed for these purposes, but challenges still exist when high accuracy is needed \cite{Cao_quantum_2019,kirkpatrick_pushing_2021}.
For instance, the sub-chemical-accuracy is often desired to predict adsorption of molecules on surfaces, the packing order of organic chemicals, and the hydrogen bonding of water and biological molecules \cite{brandenburg_interaction_2019,al2021interactions}.
Therefore, pushing the limit towards the exact ground state wavefunction of molecules is of both fundamental importance and practical relevance.  

Stochastic approaches, i.e. quantum Monte Carlo (QMC) methods, have been a competitive rival of the deterministic methods in chasing the ground truth of many body electronic wavefunction of molecules \cite{eriksen_ground_2020,simons2020,booth2009fermion,PhysRevLett.98.110201}.
In particular, diffusion Monte Carlo (DMC), an approach based on ground state projection, is capable of treating dynamic correlations and reaching sub-chemical-accuracy for molecules \cite{kent_qmcpack_2020,needs_variational_2020}.
However, effective DMC algorithms usually work together with the so-called fixed-node approximation \cite{anderson1975random,anderson1976quantum}, and the accuracy is only assured when a good trial wavefunction containing the correct nodal structure is provided in advance \cite{foulkes_quantum_2001}.
Despite many progresses have been made to improve the trial wavefunction, e.g. using physically more meaningful ansatz or
combined
with multi-determinant post Hartree-Fock wavefunctions \cite{needs_variational_2020,PhysRevB.77.115112,PhysRevE.74.066701}, the fixed-node approximation remains as the Achilles' heel of DMC.

Recently, it has been shown that machine learning techniques such as neural networks can lend strong support to describe the electronic structure of molecular systems and provide a powerful way to reconstruct the many body wavefunction \cite{han_solving_2019,ferminet,spencer_better_2020,hermann2020deep,paulinet_excited,lin2021explicitly,deeperwin,gao_pes}. 
FermiNet is one of the notable examples, which has already shown promising results for small molecules consisting of typically less than 30 electrons \cite{ferminet,spencer_better_2020,FermiNet2020github}.
In these neural network wavefunction methods, variational Monte Carlo (VMC) is often employed to train the network on the fly.
Despite its effectiveness on small molecules, it remains to be challenging to apply neural-network based VMC on larger systems, due to required large computation resources and long converging time.

In this work, we integrate the FermiNet neural network wavefunction into DMC.
This approach takes advantage of the accurate trial wavefunction of FermiNet and the efficient ground state projection of DMC,
which allows calculations of a range of systems to unprecedented accuracy. 
We refer to the vanilla FermiNet approach as FermiNet-VMC, and refer to our FermiNet-based DMC approach as FermiNet-DMC.
Compared to FermiNet-VMC, FermiNet-DMC is able to achieve lower variational ground state energy at reduced computational cost.
We carry out tests on atoms as well as molecules including $\text{N}_2$, cyclobutadiene, water dimer, benzene and benzene dimer.
We also present the empirical linear relation between VMC and DMC energies in our calculations and introduce an extrapolation scheme accordingly.
Insights to the electronic structure of these systems obtained from our calculations are also discussed.

\section{results}

\subsection{Computational framework}

\begin{figure*}[t]
    \centering
    \includegraphics{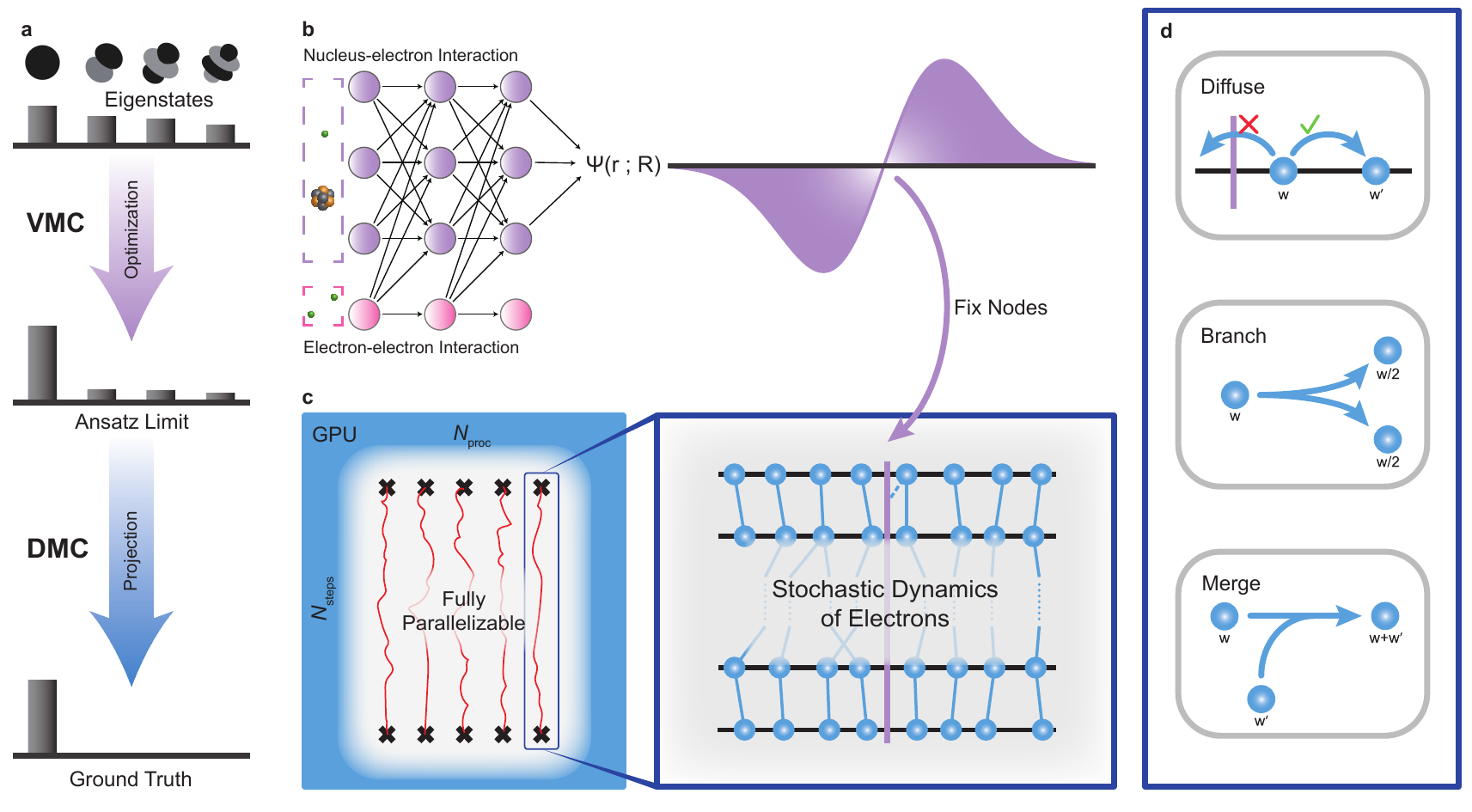}
    \caption{{\bf Computational framework.} \textbf{a}, A sketch of a brief overview on variational Monte Carlo (VMC) and diffusion Monte Carlo (DMC) from the perspective of eigenstates composition.
    Atomic orbitals represent different eigenstates, and histograms indicate the weight of each eigenstate in the state decomposition. 
    Top: A randomly initialized state with no dominant eigenstate.
    Middle: The output state of VMC where the ground state dominates, but other eigenstates are still non-negligible due to ansatz limitations.
    Bottom: The output state of DMC, which surpasses ansatz limitations and reaches the ground state.
    \textbf{b}, Left: a neural network ansatz of wavefunction; right: one dimensional projection of a many electron wavefunction and its nodal surface. 
    \textbf{c}, Left: parallelized diffusion Monte Carlo processes on GPU; 
    right: zoom in to the stochastic dynamics of each walker containing configurations of all electrons in the system, while the nodal structure is fixed. 
    This panel is inspired by and adapted with permission from the website of Quantum Monte Carlo for Chemistry @ Toulouse (\protect\url{http://qmcchem.ups-tlse.fr/index.php/Quantum\_Monte_Carlo_for_Chemistry_@_Toulouse}) \cite{dmc_figure}. 
    \textbf{d}, Three key steps in diffusion Monte Carlo. Each walker is assigned with a weight which evolves every iteration. 
    Diffuse: The stochastic propagation of walkers without crossing the nodal surface.
    Branch: Split one walker when its weight becomes too large.
    Merge: Merge two walkers when their weights become too small.
    }
    \label{framework}
\end{figure*}

As illustrated in Fig. 1a, in the traditional electronic structure approach, diffusion Monte Carlo is often used after optimization of trial wavefunction using VMC, which approaches the limit of a given wavefunction ansatz. 
DMC further purifies the true ground state out of other contaminating eigenstates, and it often allows the breaking through of the ansatz limit. 
However, to overcome the notorious sign problem, nodal points where the wavefunction is zero have to be fixed in DMC, and walkers are only allowed to evolve in each fixed nodal pocket.
Here, the idea is to implement the recently developed neural network as an accurate wavefunction ansatz (Fig. 1b).
On one hand, the wavefunction learned by the neural network automatically reproduces an accurate representation of the mysterious nodal structure of many electrons of molecules. 
The accurate nodal structure ensures that the subsequent DMC simulation with fixed nodes does not yield bias to the ground state.
On the other hand, compared with neural networks based VMC, our scheme only requires the information of the nodal structure instead of the full wavefunction.
It is reasonable to expect the nodal structure to be simpler characterized than the full wavefunction.

Our multi-walker DMC algorithm is implemented in a fully parallel manner, in which each walker independently simulates the stochastic dynamics of electrons (Fig. 1c).
The three key steps in our DMC algorithm are diffusion, branching, and merging (Fig. 1d), and they ensure that the equilibrium is reached for each walker after simulation in terms of the probability distribution of different electronic configurations.  
The diffusion step changes the configuration of electrons from one to another, while the cross-node movement is forbidden.
Branching and merging control the total population of walkers during the simulation. 
In this work, we have implemented a GPU and neural network friendly DMC algorithm, which can be easily scaled out to multiple computing nodes.
The runtime for one step in FermiNet-DMC is almost identical to that in FermiNet-VMC.
Therefore, to compare the efficiency or total runtime between FermiNet-DMC and FermiNet-VMC, we only need to compare the number of steps in those processes.
More methodological and technical details are provided in the Method section and the Supplementary Note 1-5.

\subsection{Single atoms}\label{sec:atoms}

\begin{figure*}[t]
    \centering
    \includegraphics{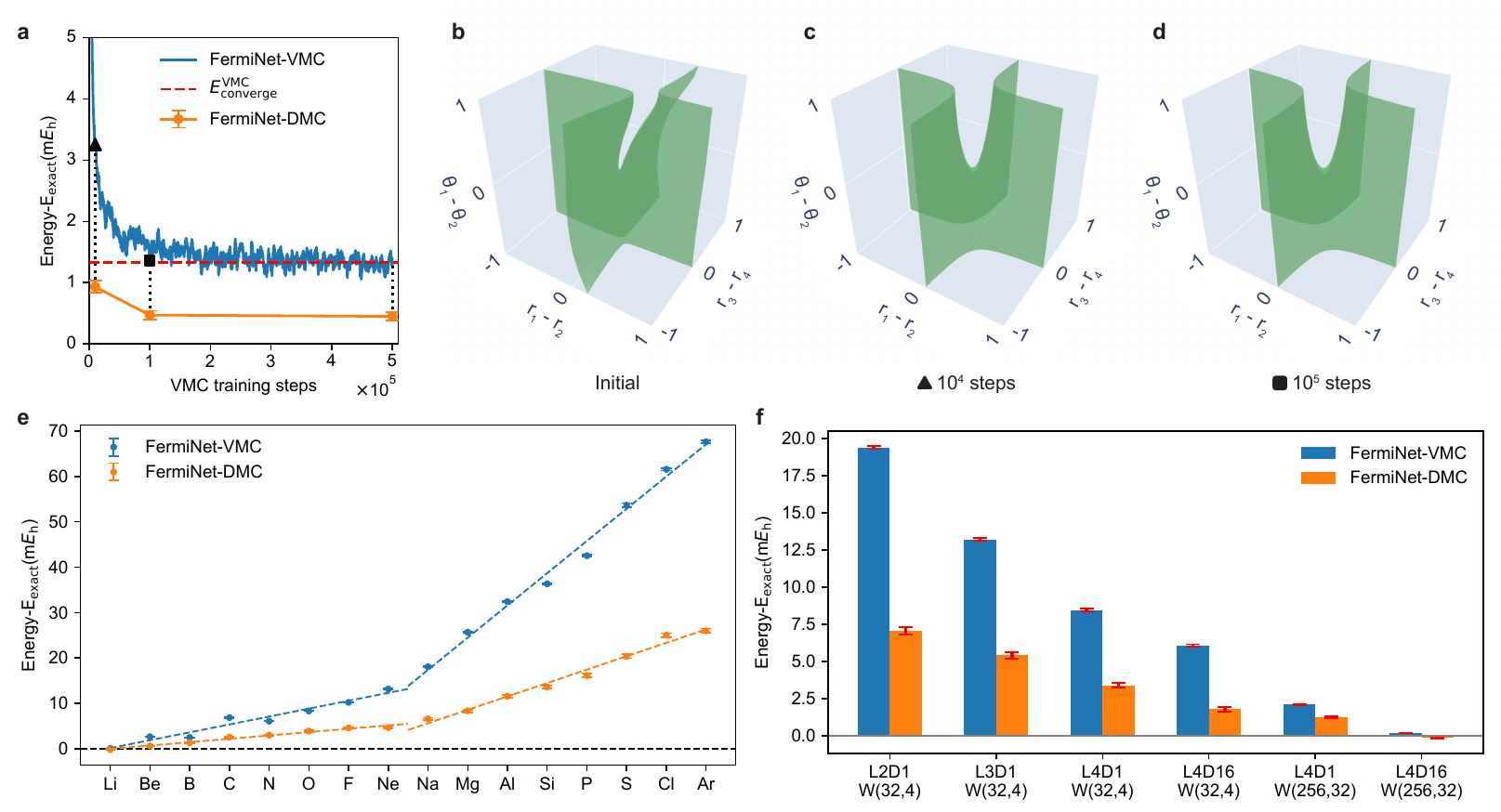}
    \caption{{\bf Accuracy and efficiency of FermiNet-DMC on single atoms.} \textbf{a}, The blue line shows the energy of a 2-layer FermiNet as a function of training iteration for a Be atom. The orange symbols show the DMC energy obtained with the trial wavefunction at the $10^4$th, $10^5$th and $5\times10^5$th training iteration, respectively. The red dashed line shows the final convergence of VMC energy. \textbf{b-d}, The three isosurfaces show the three-dimensional cuts of the full 11D nodal structure obtained at different training iterations (See Methods, section Nodal structure and wavefunction visualization for plotting details). \textbf{e}, The calculated energies of atoms with respect to the reference ground state energy as the number of electrons increases with a 2-layer network. The dashed lines are linear fittings on the second and the third period elements, respectively. \textbf{f}, The energy error with different settings of FermiNet for a Ne atom. Three variables are respectively the number of layers (L), the number of determinants (D) and the width (W) of each layer. All the energy reference values are from Chakravorty \textit{et al.} (1993)\cite{chakravorty1993ground}.}
    \label{acc_eff}
\end{figure*}

Neural network models are faced with the trade-off between model expressiveness and computational intensiveness. 
For powerful models like FermiNet, it may take hundreds of thousands iterations to converge in the training process even for small benchmark systems with just a few electrons.
Fig. 2 shows calculations for single atoms with a shallow and narrow FermiNet ansatz with only 2 layers of rather small number of neurons (see Supplementary Table S3 for details).
The network is designed to be restricted so that we can study FermiNet's performance when it is not expressive enough for the considered systems.
This situation is of practical importance especially when we are interested in applying neural network based QMC methods to large systems of one hundred electrons or more.
As shown in Fig. 2a, a common pattern of FermiNet's training progress is that the energy curve drops to a fairly low level in a short amount of time and then slowly converges to its limit. 
Fig. 2a is a calculation on the Be atom with the mentioned small network, and after $5\times10^5$ steps of training which ensures complete convergence, the systematic error still can not be reduced to within the chemical accuracy.
In addition, the computational cost could scale up quickly for larger systems even on the most advanced modern computation platforms such as NVIDIA's Tesla A100 GPU. 
This issue prevents accurate calculations for more than 30 correlated electrons \cite{ferminet,spencer_better_2020}.

The combination of the FermiNet neural network wavefunction ansatz and DMC achieves a substantial improvement in both accuracy and efficiency.
For Be atom and the same simple neural network, FermiNet-DMC energy
drops to within 1 mHa with respect to the reference value of the total energy.
The DMC data is obtained with $10^5$ steps of simulation, and the variance of DMC is also significantly reduced.
It is also encouraging to see that even when we start from the trial wavefunction after $10^4$ steps of training, the DMC energy obtained subsequently is also converged within 1 mHa to the exact value. 
At the $10^5$ step when the training has not yet completely converged, the DMC energy is already consistent with the result obtained at the $5\times10^5$ step.
The good performance of DMC based on undertrained trial wavefunctions suggests the nodal structure is well characterized before the wavefunction is fully trained in the neural network ansatz. 
In Fig. 2b-d, we present the three-dimensional cuts of the full 11-dimensional (11D) nodal structure of the FermiNet wavefunction at the initial, $10^4$, and $10^5$ step. %
The $10^5$ step nodal structure is very well converged to the correct one obtained from CI calculations \cite{bressanini2012implications}, and the nodal structure at $10^4$ step is also qualitatively same, explaining the high accuracy obtained subsequently using DMC.
For comparison, the nodal structure of the initial wavefunction is also shown.
Because of the fact that only the nodal structure determines the accuracy of DMC, the training process of neural network functions can be significantly shortened.
Overall, to reach chemical accuracy for Be atom, the cost of FermiNet-DMC is only a fraction of the cost of FermiNet-VMC.

Fig. 2e further shows the energy of FermiNet-based VMC and DMC for different atoms in order of the number of electrons under the same 2-layer network. 
We try different learning rates and train enough iterations ($10^6$ for S, Cl, Ar and $5\times10^5$ for the other atoms) to ensure that we make full use of the expressive power of the network.
As expected the error of VMC increases when the number of electrons in the system increases and the complexity of the system gradually exceeds the expressive limits of the neural network. 
With DMC the errors are reduced by more than half.
The dashed lines are linear fittings of the VMC and DMC energy. And the deviation of the data points from the fitting curves indicates that there is a correlation between the DMC and the VMC energy: when the VMC energy is comparably better, the DMC error is also smaller.
The linear rising of DMC error shows that the training of nodal structure also becomes increasingly difficult when system size increases, and the correlation between the VMC and the DMC error indicates the information of the nodal structure is closely entangled with the full wavefunction.
Note that we use a 2-layer network here in order to examine the behavior of FermiNet VMC and DMC in the regime where the network ansatz is relatively restricted for the considered systems, while FermiNet-VMC can be more expressive to achieve high accuracy for those atoms with more layers and neurons, as shown in \cite{ferminet}.

Moreover, the improvement of DMC suggests that it may take a smaller and hence more efficient network to represent the nodal surface, without affecting the DMC accuracy.
In Fig. 2f, we present a set of such tests on Ne atom, where the complexity of the neural network is labeled as (L,D,W) to indicate the number of layers, the number of determinants, and the width of each layer in the network, respectively.
Overall, when the expressiveness of the network is reduced both VMC and DMC are affected in terms of their accuracy.
Therefore, all the calculations suggest that the VMC energy is a good indicator of not only how well the wavefunction is optimized but also the quality of its nodal structure.
The behavior is also expected for other neural network wavefunction ansatz.
Combined with the typical first-steep-then-flat optimization curve in neural networks, we can automate the switching-on of DMC and minimize the total cost of calculations at targeted accuracy.

Building upon the successful treatments of FermiNet-DMC on atoms, we now extend the approach to larger molecules.

\begin{figure*}[t]
    \centering
    \includegraphics[width=174mm]{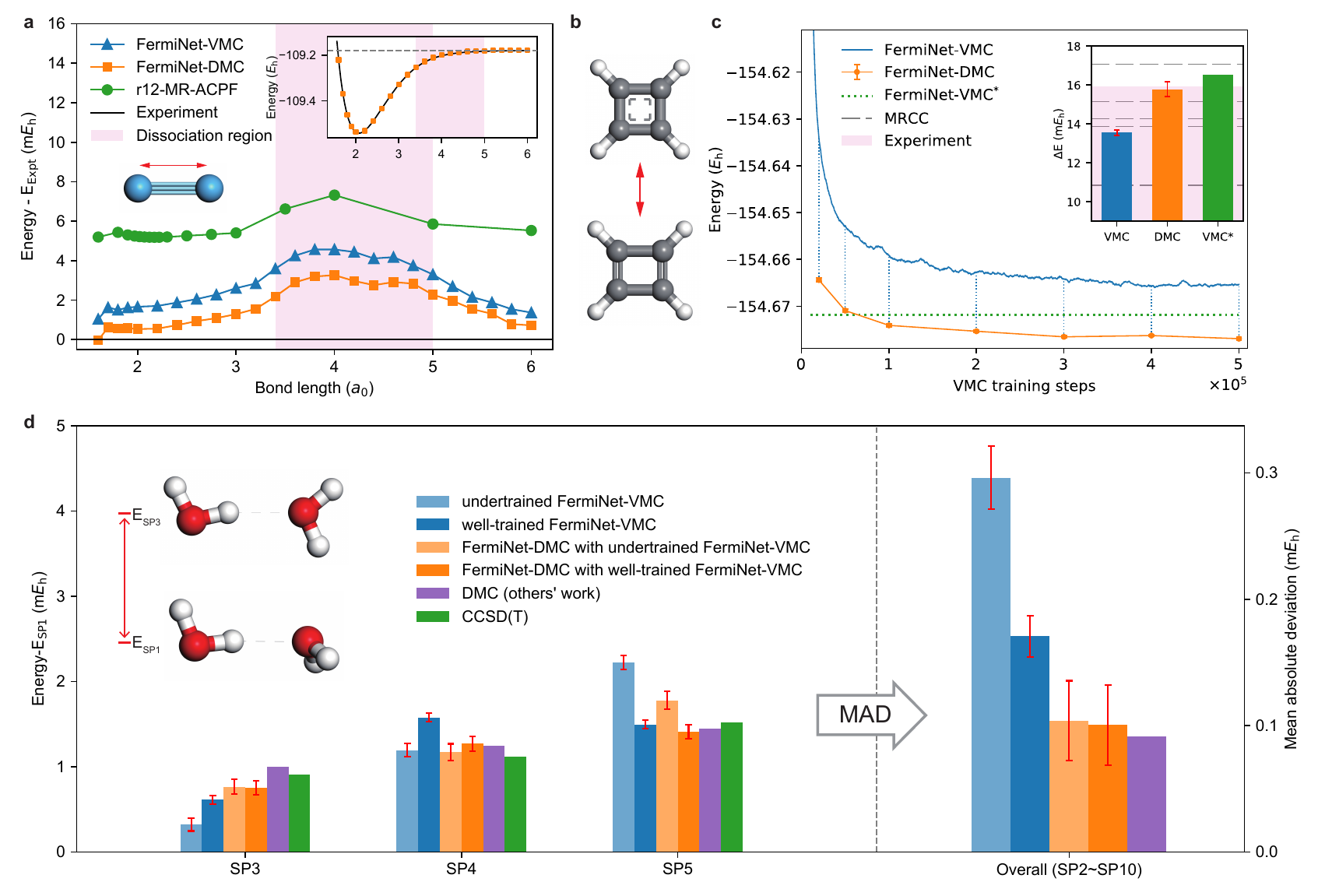}
    \caption{{\bf Calculations on N$_2$, cyclobutadiene and water dimer.} \textbf{a}, Main panel: calculated energy of N$_2$ at different bond length, plotted as the difference to the experimental data \cite{le2006accurate}. For comparison, the green line is the highly accurate r12-MR-ACPF results under a modified basis set based on aug-cc-pV5Z \cite{gdanitz1998accurately}. Inset: the dissociation curves from experimental data (black line) and FermiNet-DMC (orange squares). The negligible error bars (less than 0.1 mHa) are not plotted. The pink backgrounds highlight the dissociation region where correlations are strong. \textbf{b}, Molecular structures of cyclobutadiene's equilibrium state (bottom) and transition state (top). \textbf{c}, Main panel: the ground state energy of cyclobutadiene's equilibrium state as a function of the VMC training step. FermiNet-DMC energy is calculated using the trial wavefunction at the corresponding training steps. 
    FermiNet-${\rm VMC}^*$ indicates the result from Spencer et al. \cite{spencer_better_2020}. 
    Inset: the transition barrier of cyclobutadiene calculated with different methods. 
    Pink background indicates the range of experimental estimates between 2.5 and 15.9 mHa, while we show only the part above 9 mHa to highlight differences between QMC results.
   Grey dashed lines indicate results from five multi-reference coupled cluster (MRCC) methods (top to bottom):  MR-DI-EOMCCSD, RMRCCSD(T), Mk-MRCCSD(T), MRCISD+Q and BW-MRCCSD(T) \cite{lyakh2012multireference}. 
   \textbf{d}, Left: three of the relative energies of the 10 Smith stationary points SP$n$ $(n=1,2,...,10)$\cite{smith1990transition}. For the other results, see Supplementary Fig. 6. The SP1 structure is the global minimum and is taken as the reference. The geometries of SP1 and SP3 are shown as insets while the others are included in Supplementary Fig. 4. Two neural networks have been trained for $10^5$ steps and $3\times10^5$ steps, which are dubbed as ``undertrained'' and ``well-trained''. All the geometries are optimized by CCSD(T) \cite{tschumper2002anchoring}. The CCSD(T) energies and the DMC results with the conventional Slater-Jastrow ansatz \cite{gillan2012assessing} are also plotted for comparison. The error bar of the energy difference is calculated as the square root of the sum of the squares of each energy estimator's standard error. Right: mean absolute deviation from CCSD(T) results over all the relative energies.}
    \label{molecules}
\end{figure*} 

\subsection{Nitrogen molecule}
The first example is the dissociation curve of $\text{N}_2$ molecule. At equilibrium $\text{N}_2$ forms a strong triple covalent bond at 2.1 a.u., and the dissociation is accompanied by a severe bond breaking process, which is strongly correlated in nature.
Therefore, the dissociation curve of $\text{N}_2$ is often used to benchmark electronic structure methods' description of strong correlation.
In DMC, this is also highly relevant because the nodal structure is directly affected by electron correlation. 
Fig. 3a plots the relative energy of $\text{N}_2$ with respect to the experimental reference \cite{le2006accurate} as a function of the bond length.
The results from FermiNet-VMC and r12-MR-ACPF, a state-of-the-art traditional multi-reference approach \cite{gdanitz1998accurately}, are also shown. 
We can see that our DMC calculations are consistently better than those references, with an error of less than 1 mHa in a wide range of bond length. 
The largest error comes, not surprisingly, around the dissociation point near 4 a.u., and yet the error is only 3 mHa.
In fact, our results can be considered as the most accurate \textit{ab initio} one of $\text{N}_2$ dissociation curve reported so far. 
It is worth noting that the FermiNet-VMC results here have been remarkably accurate, whose deviation from experiment curve is within 2 mHa near equilibrium and 4 mHa in dissociation region.
Yet our FermiNet-DMC results can improve averagely about 1 mHa.
For comparison, CCSD(T) calculation (not plotted), which is known as the ``golden standard'' in quantum chemistry, have an error of 25 mHa around 4 a.u. \cite{ferminet}.
In terms of relative energy, the non-parallelity error (NPE) of FermiNet-DMC (3.28 mHa) is only slightly better than that of FermiNet-VMC (3.53 mHa), consistent with mild improvement on small systems reported in \cite{dmcGao}, and both are comparable to the state-of-the-art r12-MR-ACPF result (2.14 mHa).

The remaining error source of DMC is the nodal structure error produced in the training of neural network using VMC, which is fully reflected on the shape of the FermiNet VMC and DMC curves. 
The results of FermiNet-DMC are close to the experimental fitting curve within 1 mHa outside the dissociation region and cannot go any lower due to the variational property.
So, when combined with a more expressive or better trained neural network that can handle the dissociation region, it is very likely that the full dissociation curve of $\text{N}_2$ can be reproduced by DMC within an error of 1 mHa, meaning that DMC can also solve strongly correlated systems within chemical accuracy.

\subsection{Cyclobutadiene}
A similar example is the structural transition of cyclobutadiene, which is also well known for its multi-referential nature. 
The neural network based VMC models \cite{hermann2020deep, spencer_better_2020} have already shown promising results on cyclobutadiene.
FermiNet-DMC can handle this system with higher accuracy and reduced computational cost.

In our experiments, VMC process takes around $3\times 10^5$ steps to converge, though the converged result is still around 7 mHa higher than the reported value in ref. \cite{spencer_better_2020}, which converges in $2\times 10^5$ steps.
This is probably because we use different training hyperparamters, or simply because our optimization process gets trapped in a bad local minimum.
However, our final DMC result is around 4 mHa lower than the reference data  \cite{spencer_better_2020}.
This demonstrates the effectiveness of our DMC implementation as a seamless extension to VMC.
Namely even if the optimization in VMC does not work well, the following DMC process can still bring the energy calculation to a highly accurate level.
This is especially important for neural-network based VMC, because its optimization is significantly trickier to tune and requires a longer time to completely converge, compared to conventional VMC.
Here the DMC finite time step error is negligible as illustrated in Supplementary Fig. 1, which guarantees the variational property of our FermiNet-DMC results,

With $10^5$ VMC and $10^5$ DMC steps, FermiNet-DMC's energy result is 2 mHa lower than the reference data in ref.~\cite{spencer_better_2020} produced from a training phase with $2\times 10^5$ VMC steps.
Note that in this case our number of total QMC steps is still slightly less than the ref.~\cite{spencer_better_2020} due to the required inference phase in FermiNet-VMC.
Therefore, FermiNet-DMC should be preferred for its lower variational energy at the same or less computational cost.

The automerization energy difference of cyclobutadiene is shown in the inset panel of Fig. 3c. 
Neural-network based VMC gives an accurate automerization energy difference of cyclobutadiene \cite{hermann2020deep, spencer_better_2020}.
It is consistent with the high-end of the experimental data. 
The results of FermiNet-DMC are also in the same region.
See Supplementary Note 9 for more details, including the training curve for transition configuration and the DMC energy data for both equilibrium and transition configurations.

\subsection{Water dimer}
In addition to the strong covalent bonding, where static correlation is more essential, molecular systems with weaker hydrogen bonding and non-covalent interactions can also be challenging because of dynamic correlations.
To this end, we have carried out FermiNet-DMC calculations on the 10 Smith stationary point of water dimer \cite{smith1990transition}.
The 10 structures, as illustrated in Fig. 3d and Supplementary Fig. 4, have different hydrogen bonding configurations and their relative energies are used to benchmark the performance of electronic structure methods and force field models on hydrogen bonding systems \cite{gillan_perspective_2016}.
With 10 total energy results (plotted in Supplementary Fig. 5) and 9 relative energy results (plotted in Supplementary Fig. 6), we can thus have a rather credible investigation on the error cancellation performance of FermiNet-VMC and FermiNet-DMC.
We compare the energy results of FermiNet-DMC with an undertrained network and a well-trained network as trial wavefunctions respectively.
The undertrained network is trained by VMC in $10^5$ steps, while the well-trained network is trained by VMC in $3\times 10^5$ steps.
CCSD(T) results \cite{tschumper2002anchoring} are displayed as benchmarks for their high accuracy for such type of systems. 

As shown in Fig. 3d, the undertrained FermiNet-VMC performs badly on SP3 and SP5, and so does the well-trained FermiNet-VMC on SP4, though some of the FermiNet-VMC results are quite close to the benchmark results (e.g. SP7 and SP8 in Supplementary Fig. 6).
On the other hand, FermiNet-DMC performs consistently well no matter which network is used as trial wavefunction, undertrained or well-trained.
Overall, the mean absolute deviations from the benchmark CCSD(T) results are also given in Fig. 3d, from which we can clearly tell the improvement of FermiNet-DMC on relative energy calculations.
For comparison purpose, we also show DMC results with traditional Slater-Jastrow wavefunction ansatz \cite{gillan2012assessing}, whose accuracy is at the same level with FermiNet-DMC as the difference is negligible compared to the statistical error.
The inferior performance of FermiNet-VMC may be due to the different degree of convergence in different systems, while FermiNet-DMC provides a more efficient and practical solution than fully converged FermiNet-VMC.

\subsection{Benzene}\label{sec:benzene}

\begin{figure*}[t]
    \centering
    \includegraphics{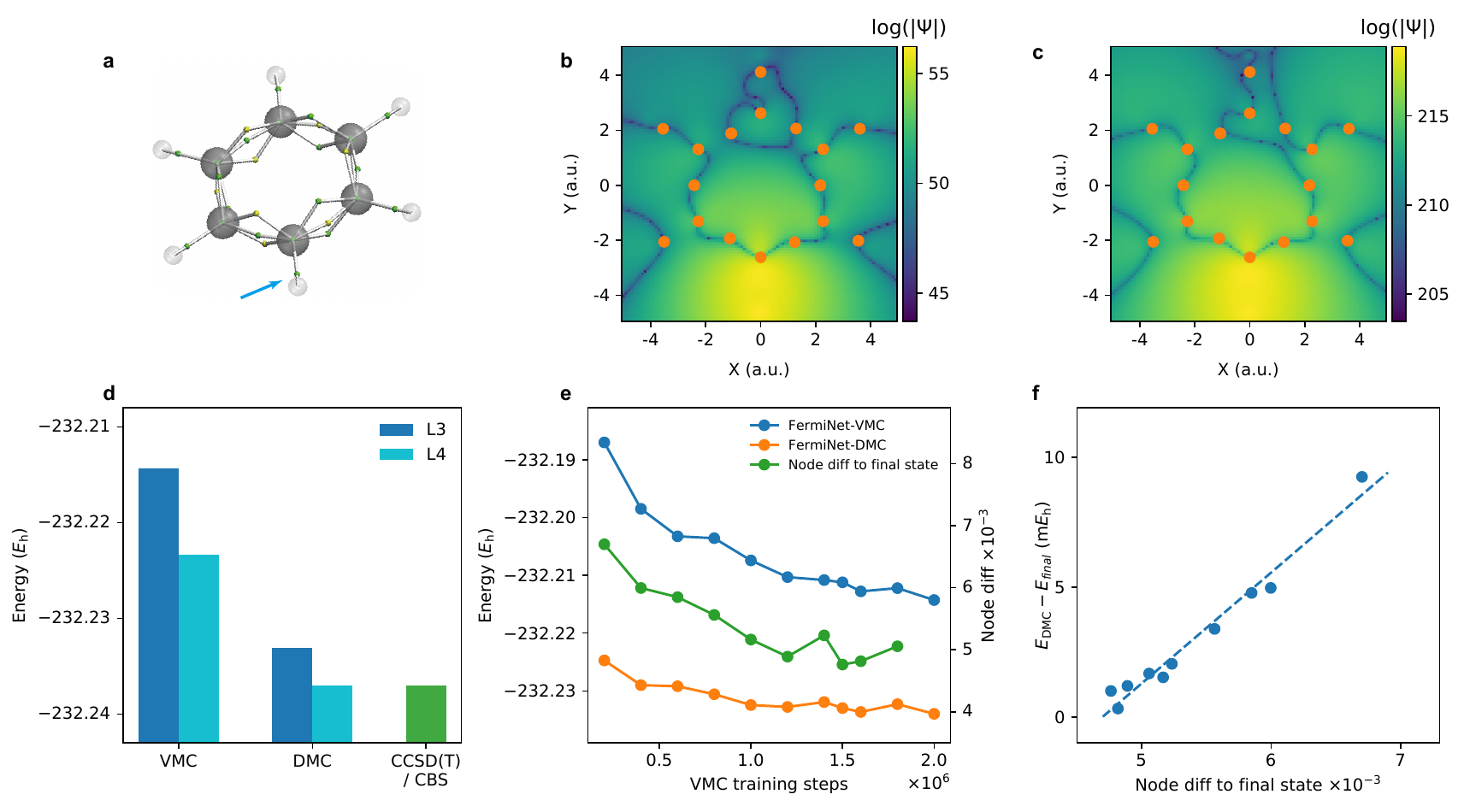}
    \caption{{\bf Calculations on benzene.} \textbf{a}, Atomic structure of the benzene molecule and representative electron positions \cite{Liu2020}. 
    Small balls represent electrons, with spin indicated by their colors, and larger balls represent nuclei. 
    The electron pointed by the blue arrow is an arbitrarily chosen one for nodal set visualization in \textbf{b} and \textbf{c}. 
    Rods are drawn to connect the nucleus with electrons nearby.
    \textbf{b} and \textbf{c}, The log-scaled magnitude of unnormalized FermiNet wavefunctions for a benzene molecule. 
    Each slice is generated by moving a single spin-up electron in the square $[-5~{\rm a.u.}, 5~{\rm a.u.}]^2$ on X-Y plane while fixing all other electrons in the representative positions shown in \textbf{a}. 
     The dark curves are the nodes and the orange points are the fixed spin-up electrons projected onto X-Y plane. 
     The moving electron for \textbf{b} and \textbf{c} corresponds to the one pointed by the blue arrow on the bottom C-H bond in \textbf{a}. 
     \textbf{b} shows a slice for a 4-layer FermiNet while \textbf{c} is for a 3-layer FermiNet.
    \textbf{d}, Ground state energy of benzene molecule. ``L3'' and ``L4'' stand for neural networks with 3 and 4 layers, respectively. The CCSD(T) result coincides with our best DMC result with the 4-layer network. 
    \textbf{e}, the trend of node difference to final state in the training process together with the ones for VMC and DMC energy for a benzene molecule using a 3-layer FermiNet.
    \textbf{f}, the linear trend between the node difference and the DMC energy difference to final state.
    The points correspond to different intermediate training steps and the dashed line is fitted using least square.
    }
    \label{fig:benzene}
\end{figure*}

To further illustrate the power of our approach, we have examined the benzene molecule and a benzene dimer.
Benzene is one of the most fundamental organic molecules with a hexagonal ring of C-H (Fig. 4a). 
There have been challenges in understanding its electronic configuration, bonding order and obtaining the ground state energy. 
To understand the electronic structure of benzene molecule, we performed FermiNet based VMC and DMC simulations with 3-layer and 4-layer networks separately. 
Our best FermiNet-DMC result calculated with the 4-layer network coincides with the CCSD(T) result extrapolated to complete-basis-set (CBS) limit.
The comparison is shown in Fig. \ref{fig:benzene}d.
The CCSD(T) result is carried out with Psi4 \cite{psi4} and the CBS result is extrapolated using cc-pCVXZ (X=3,4,5) basis, which is much larger than the ones reported in \cite{NIST} and used by others as the state-of-the-art electronic structure methods in Refs. \cite{Eriksen2020}.
The energy from our CCSD(T) / CBS calculation is also much lower than those references.
See Supplementary Note 14 for more details on the CCSD(T) calculation and CBS extrapolation.

The 3-layer FermiNet here is much smaller than the 4-layer one.
Besides being one layer shallower, the number of neurons on each layer is also significantly less. 
See Supplementary Tables 8-10 for the related hyperparameters.
Fig. 4d shows that the 3-layer FermiNet-DMC's energy is lower than the 4-layer VMC result by around 10 mHa, which demonstrates one of the main benefits of FermiNet-DMC that it can achieve better accuracy with smaller network.
This is especially important when we are dealing with large systems.

In our calculations, FermiNet-DMC is able to achieve lower variational energy results with an order of magnitude better efficiency.
With a total of $4\times 10^5$ QMC steps ($2\times 10^5$ VMC training steps and $2\times 10^5$ DMC steps), the 3-layer FermiNet-DMC's energy result (-232.225 Ha) is slightly better than the 4-layer FermiNet-VMC's energy result (-232.223 Ha) at $10^6$ VMC training step. 
Moreover, the runtime of a single VMC step for the 4-layer network is approximately 4 times that of a single VMC or DMC step for the 3-layer network under the same computation resources.
Therefore, in this case, the 3-layer FermiNet-DMC can achieve a better energy result at only a tenth of the total computation cost compared to the 4-layer FermiNet-VMC.
Similarly, compared to the 3-layer FermiNet-VMC at $2\times 10^6$ VMC training step, the 3-layer FermiNet-DMC with $4\times 10^5$ QMC steps can achieve more than 10 mHa better energy result at only a fifth of the total computation cost.

Furthermore, the energy difference between the FermiNet-DMC results in Fig. 4d is only around 3 to 4 mHa, suggesting the closeness between the node structure of the two trial wavefunctions. 
To confirm this statement, we visualized 2-dimensional slices of those trial wavefunctions in Fig. \ref{fig:benzene}b-c. 
The slices are generated by moving a single spin-up electron inside a two dimensional box while fixing all other electrons at representative positions suggested by Liu et al. \cite{Liu2020} and illustrated in Fig. \ref{fig:benzene}a.
See Section \ref{sec:method_visualization} and Supplementary Note 11 for more visualization details.
Comparing Fig. 4b (4-layer FermiNet VMC) and Fig. 4c (3-layer FermiNet), we find that the nodes, represented by the dark pixels, do share the same pattern.
Moreover, the parts of nodal surface in lighter areas, namely with larger wavefunction value, are very close to each other in Fig. \ref{fig:benzene}b-c, and they are the most important parts of nodal surface in the DMC process since walkers are more likely to visit its neighbourhood.
The closeness of those parts is consistent with the fact that the FermiNet-DMC energies are close.

To track how nodal surface evolves along the training process, we propose a divergence  $D(S, T)$ measuring the difference between two nodal surfaces $S$ and $T$.
The definition and algorithmic details are described in section \ref{sec:node_diff} and Supplementary Note 15, and the definition is also related to the intuition mentioned above that nodes in the neighbourhood with larger wavefunction value are more important in the QMC calculation.
For the 3-layer FermiNet, we calculated 
$$D(S_{\rm final}, S_{k}),$$
where $S_{\rm final}$ and $S_k$ are the nodal surface corresponding to the final VMC training step and  the intermediate training steps $k$, respectively.
The result is shown in Fig. \ref{fig:benzene}e together with the VMC and DMC energy, where the trend of the divergence correlates well with energies.
As a matter of fact, there is a linear relation between the divergence and DMC energy, as shown in Fig. \ref{fig:benzene}f, indicating that the proposed divergence successfully captures the essential information of the difference between nodal surfaces.
Here the divergence converges to around 0.005 instead of 0 because of the large learning rate used when training the 3-layer FermiNet for benzene.

\begin{figure*}[t]
    \centering
    \includegraphics{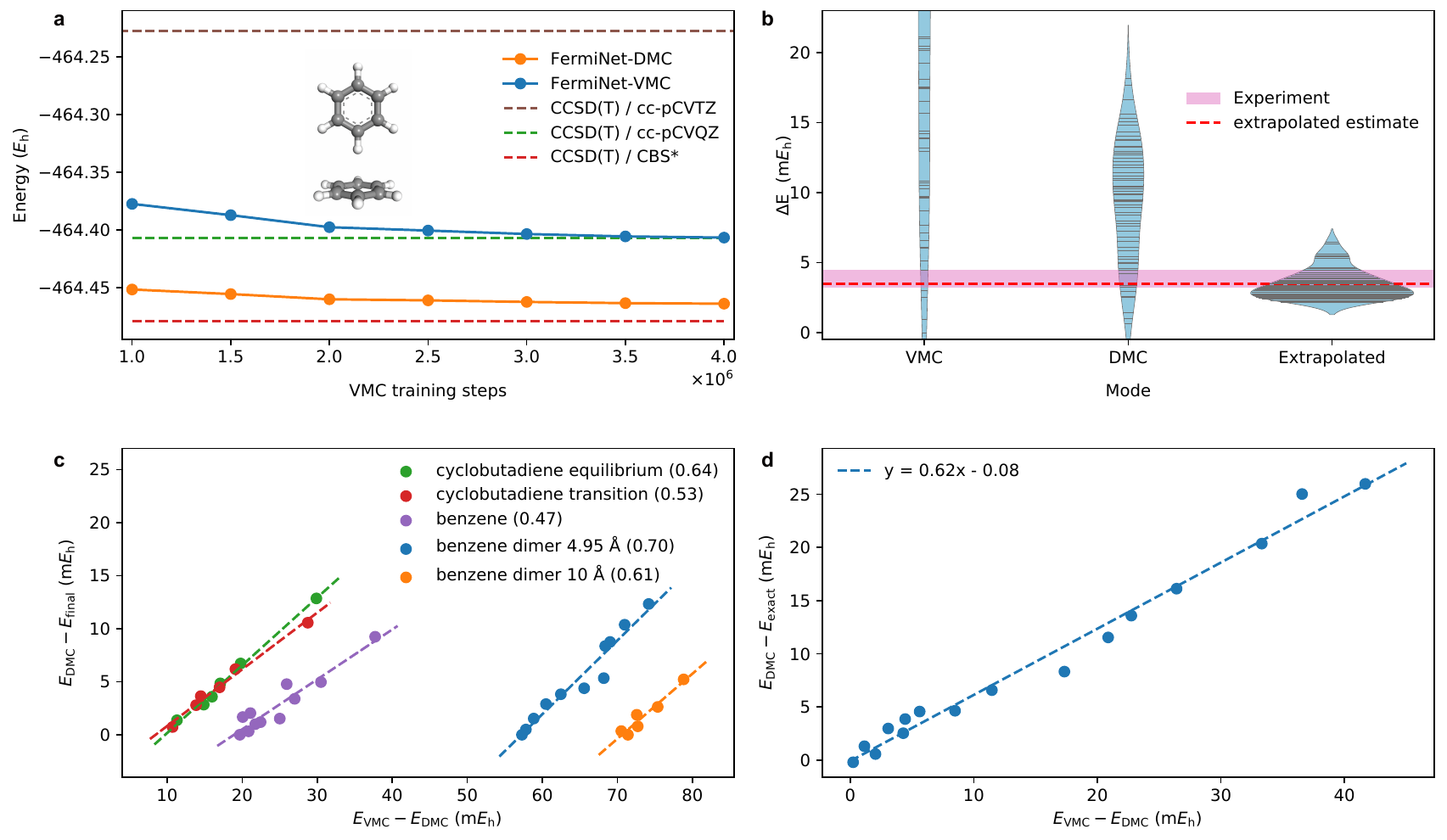}
    \caption{{\bf Benzene dimer calculated energy and the extrapolation based on VMC-DMC linear relation.}
    \textbf{a}, The energy of the T-shaped benzene dimer with a center-to-center distance of 4.95 \AA~as well as the CCSD(T) results as baselines. 
    The CCSD(T) / CBS* result is calculated using binding energy and monomer energy. 
    See Supplementary Note 14 for details.
    \textbf{b},  The fitted distribution of binding energy from VMC, DMC and extrapolation results, where the solid lines inside each violin represent the actual data points.
    The energy for two separated benzene molecules is calculated using a benzene dimer separated by 10 \AA~\cite{azadi2015chemical}. 
    The equilibrium configuration is a T-shaped structure with center-to-center distance of 4.95 \AA~\cite{pitonak2008benzene}.
    The experimental range is from Grover et al. \cite{bd_exp}. 
    The extrapolation scheme is based on the empirical linear relation between VMC energy ($E_{\rm VMC}$) and DMC energy ($E_{\rm DMC}$) in the training process, indicated by the blue and yellow points in \textbf{c}.
    \textbf{c}, Linear fitting of $E_{\rm VMC} - E_{\rm DMC}$ with respect to $E_{\rm DMC} - E_{\rm final}$ on various molecular systems, where $E_{\rm final}$ represents the DMC energy at final VMC training step. 
    The number within the parenthesis in the legend is the fitted slope for each system.
    \textbf{d}, Linear fitting of $E_{\rm VMC} - E_{\rm DMC}$ with respect to $E_{\rm DMC} - E_{\rm exact}$ on atoms, where the energy data are the same as Fig. \ref{acc_eff}e.
    Note that the data in \textbf{d} is fundamentally different from \textbf{c} in the sense that the linearity in \textbf{c} is about different training steps for the same system while in \textbf{d} the linearity is measured across different systems.
    }
    \label{fig:linearity}
\end{figure*}

We have also trained a neural network for a benzene dimer, which is a prototypical system to further test non-covalent interactions. 
Benzene dimer, which has 84 electrons in total, is a much larger system than the ones considered in previous neural-network based VMC works \cite{han_solving_2019,ferminet,spencer_better_2020, hermann2020deep,lin2021explicitly, deeperwin,gao_pes,paulinet_excited}.
We elaborate the challenges and tricks dealing with large systems using FermiNet based QMC methods in Supplementary Note 5.
We consider a T-shaped structure with an edge-to-face arrangement, as illustrated in Fig. \ref{fig:linearity}a, specifically the equilibrium configuration with a center-to-center distance of 4.95 \AA~\cite{pitonak2008benzene}.
Fig. \ref{fig:linearity}a also shows the VMC and DMC energies as functions of the VMC training step, which are both over 200 mHa lower than the CCSD(T) result with cc-pCVTZ basis.
The converged FermiNet-DMC energy is over 50 mHa lower than both FermiNet-VMC result and the CCSD(T) result with cc-pCVQZ basis. 
It echos statements made in above sections that FermiNet-DMC can achieve significantly higher accuracy for larger systems or cases where the neural network ansatz is not powerful enough to characterize the ground state wavefunction well.
For comparison, the FermiNet-VMC energy has not fully converged even after four million training steps. 
Sch{\"a}tzle et al. \cite{schatzle2021convergence} shows that neural-network based VMC, in particular, PauliNet, can achieve variational energies at the fixed-node limit in certain circumstances, while in our calculations, one can clearly see that it is not the case for FermiNet especially when its expressive power is limited compared to the size of the system.
On the other hand, our DMC result is 15 mHa higher than the CCSD(T) / CBS result.
Note that CCSD(T) is not a variational method, hence the relatively lower CCSD(T) / CBS result may indicate similar accuracy compared to our DMC result.
To achieve more accurate FermiNet-DMC result, we can use a better neural-network trial wavefunction with a larger network or a better network architecture.

In addition to the total energy at the equilibrium configuration, binding energy is also of great interest when studying a benzene dimer \cite{pitonak2008benzene,Al-Hamdani2021,azadi2015chemical,Tsuzuki_benzene_dimer,Sinnokrot_benzene_dimer}, and classical methods, such as CCSD(T) and MP2, can produce results agreeing with experimental data well.
However, for neural-network based QMC, the binding energy calculation is more subtle and challenging due to the lack of systematic error cancellation. 
Using the same network structure handling both monomer and dimer would introduce additional size-inconsistency-like bias because of the more severe expressiveness limitation on benzene dimer than monomer.
For the benzene dimer, we find such an estimate would predict a severe underbinding with both VMC and DMC.
Another way to estimate the binding energy is to take the difference between a separated dimer configuration (10 \AA) \cite{azadi2015chemical} and the equilibrium configuration, shown in Fig. \ref{fig:linearity}b, which turns out to be systematically overbinding.
With an empirical linear relation between VMC and DMC energy in the training process, we developed a simple VMC-DMC hybrid extrapolation scheme, which leads to an accurate estimate of the binding energy, well-agreed with the experimental measurements \cite{bd_exp}, also displayed in Fig. \ref{fig:linearity}b. 
We will elaborate more on this extrapolation scheme in Section \ref{sec:linearity}.
In order to systematically improve the binding energy calculation, the most straightforward way is to adopt better neural-network ansatz as the trial wavefunction for better accuracy.
Adding regularization mechanism in the optimization processes is another possible option so that the model variance can be reduced for better error cancellation.
Note that in the case of DMC with pseudopotential, binding energy calculation can be also improved with certain deterministic approximation \cite{zen2019new}. 
We will leave it as a future study apply those ideas to improve the binding energy calculation.

\subsection{Linear relation between VMC-DMC energy}\label{sec:linearity}
Quite consistently, we find linear relation between VMC and DMC energies in our calculation.
We have encountered two types of linear relation.
One is about intermediate energies calculated along the training process for a given system, while another one is about the converged energies from different systems. 
We take advantage of the first type of linearity and develop a simple but effective extrapolation scheme accordingly.

We find that, for molecular systems, such as cyclobutadiene, benzene monomer and dimers, there's a linear trend between the VMC and DMC energies calculated at different steps along the VMC training process. 
Equivalently, there's a linear relation between quantities 
\begin{equation*}
    E_{\rm DMC}^{(k)} - E_{\rm final} \text{ v.s. } E_{\rm VMC}^{(k)} - E_{\rm DMC}^{(k)}
\end{equation*}
where $E_{\rm VMC}^{(k)}$ and $E_{\rm DMC}^{(k)}$ represent the VMC and DMC energy calculated at VMC training step $k$, and $E_{\rm final}$ is the DMC energy at the final VMC training step, namely a constant for one training process. 
Such relation is shown in Fig. \ref{fig:linearity}c.
Based on this empirical linear relation, we propose an extrapolation scheme 
\begin{equation}
    E_{\rm DMC}^{(k)} - E_{\rm ex} = w\cdot (E_{\rm VMC}^{(k)} - E_{\rm DMC}^{(k)}) + b
\end{equation}
where $E_{\rm ex}$ is the extrapolated energy, and $w$ and $b$ are two parameters to be determined. 
Here slope $w$ can be fitted using $E_{\rm VMC}^{(k)}$ and $E_{\rm DMC}^{(k)}$ along the training process, but the intercept $b$ cannot be inferred from those data.
Therefore it is difficult to use this scheme to extrapolate absolute energy unless we have extra information on intercept $b$.
On the other hand, when calculating relative energy, we may simply assume the intercept $b$ between different configurations are the same so that it can be canceled out in the calculation.
Namely for relative energy, we have
\begin{equation}
    \Delta E_{\rm ex} = (1 + w) \cdot \Delta E_{\rm DMC} - w \cdot \Delta E_{\rm VMC}
\end{equation}

Note that the calculation of relative energy is especially troublesome for neural-network based QMC methods, due to the strong dependence on the number of training steps and the long converging period.
See Supplementary Figure 8b for how binding energies calculated with FermiNet VMC and DMC change along the optimization process.
With our scheme, the binding energy results calculated from different VMC training steps would be the same, modulo the fitting error of the linear relation, which means we can circumvent the dependence of the binding energy result on the number of training steps.
In practice, the extrapolated binding energies form a well concentrated distribution, and doing an extra average using different VMC training steps can eliminate the linear fitting error and provide an accurate estimate.
Moreover, it also suggests that we can calculate the extrapolated binding energy with data collected in the early phase of the training process, avoiding the long converging period of VMC optimization.

Applying this scheme to binding energy calculation of a benzene dimer, the result is significantly improved and the distribution fitted from energy difference of different VMC training steps is concentrated around the experimental range, as shown in Fig. \ref{fig:linearity}b.
The estimate of extrapolated binding energy by averaging the energy difference is 3.60 mHa, within the experimental range.
See Supplementary Note 13 for more extrapolation-related details for benzene dimer.

We have discussed the relation of VMC and DMC energy for elements on the second and third rows in Section \ref{sec:atoms}.
For each atom, we have a reference energy data $E_{\rm exact}$ to be compared with converged VMC energy ($E_{\rm VMC}$) and DMC energy ($E_{\rm DMC}$).
As shown in Fig. \ref{acc_eff}e, both $E_{\rm DMC} - E_{\rm exact}$ and $E_{\rm VMC} - E_{\rm exact}$ grow linearly as the atomic number increases, though the slope changes when switching from the second row elements to the third row.
However, if we instead compare $$E_{\rm DMC} - E_{\rm exact} \text{ v.s. } E_{\rm VMC} - E_{\rm DMC},$$ then we have a single linear relation across all elements on both second and third rows, as shown in Fig. \ref{fig:linearity}d.

Interestingly, the slope of fitted lines in both Fig. \ref{fig:linearity}c and \ref{fig:linearity}d are all quite close. 
We will leave further study on those two types of linearity as future work.

\section{discussion}

FermiNet-DMC is able to achieve accurate \textit{ab initio} calculations for various systems, obtaining ground state of 16 atoms, N$_2$ along the bonding curve, 2 cyclobutadiene configurations, 10 hydrogen bonded water dimers, benzene monomer and dimer. 
These systems include 
bond breaking structures where strong static correlation exists and weakly bonded dimers where dynamic correlation dominates, and FermiNet-DMC performs consistently well. 
FermiNet-DMC leverages the expressive power of neural network to provide well-behaved trial wavefunctions.
Neural network based VMC has claimed success in small systems when the network can be sufficiently trained. 
However, it is not able to provide satisfactory ground state wavefunction and energy when the expressiveness of the neural network is limited.
Compared to VMC, the combination of neural network with DMC provides a powerful solution, in the sense that it can achieve more accurate result with simpler network and better efficiency. 
The improvement of FermiNet-DMC in efficiency can be up to 1 or 2 orders of magnitude in the large systems tested in order to reach the same accuracy level as FermiNet-VMC, which can become increasingly more important when dealing with even larger molecules.

There is an interesting linear relation between VMC and DMC energy observed during the training process as well as across different systems. 
We develop an extrapolation scheme accordingly, which greatly improves the accuracy of relative energy calculation as shown in the benzene dimer case and overcome the issue that the relative energy calculation greatly depends on the different training steps in the QMC process. 
We also design a divergence measuring the difference between nodal surfaces of two wavefunctions, which correlates well with the corresponding DMC energies in numerical experiments.
Namely the proposed divergence successfully captures the essence of nodal surface differences.

It is worth pointing out that a similar idea to this work was proposed in a preprint by Wilson et al., where they have performed preliminary tests on the second row elements \cite{dmcGao}. 
However,
only minor improvements in accuracy were observed
accompanied by an increased cost of DMC, since the FermiNet used there was powerful enough to achieve high accuracy for the tested small systems and leave little room for further improvement.
By comparison, our approach, being more sophisticated and efficient, achieves significant accuracy boost when dealing with more challenging molecular systems, which FermiNet alone cannot handle well.
We have also shown that even for small systems, FermiNet-DMC should still be preferred for the fact that it can achieve comparable or even better accuracy with a smaller network and much less computation resources compared with FermiNet-VMC.
Our work, therefore, eliminates the negative concerns of going from VMC to DMC with neural network wavefunction ansatz. 
Moreover, the DMC method can be further integrated with other powerful molecular neural networks \cite{hermann2020deep,deeperwin}, periodic neural network for solids \cite{li_ab_2022}, neural networks with effective core potential \cite{PhysRevResearch.4.013021}, which has the potential to catalyze a paradigm shift in the application of stochastic electronic structure methods.

\section{methods}
\subsection{Basic theory}
To study a many-body system from first principles, we always consider solving the well-known Schr\"{o}dinger equation for electrons and nuclei.
When we work in the Born-Oppenheimer approximation \cite{born1927quantentheorie}, and further consider a fixed set of nuclear positions, the problem is simplified to the solution of the ground state many-electron wavefunction. 
\begin{eqnarray}
\hat{H}\psi&&(\bm{\mathrm{x}}_1,\cdots,\bm{\mathrm{x}}_n)=E\psi(\bm{\mathrm{x}}_1,\cdots,\bm{\mathrm{x}}_n),\\
\hat{H}=&&-\frac{1}{2}\sum_i\nabla_i^2-\sum_I\sum_i\frac{Z_I}{\left|\bm{\mathrm{r}}_i-\bm{\mathrm{R}}_I\right|}\nonumber\\
&&+\sum_{i<j}\frac{1}{\left|\bm{\mathrm{r}}_i-\bm{\mathrm{r}}_j\right|}+\sum_{I<J}\frac{Z_I Z_J}{\left|\bm{\mathrm{R}}_I-\bm{\mathrm{R}}_J\right|},\nonumber
\end{eqnarray}
where $\bm{\mathrm{x}}_i=(\bm{\mathrm{r}}_i, \sigma_i)$ denotes the spatial and spin coordinates of electron $i$, and $\bm{\mathrm{R}}_I, Z_I$ respectively denote the spatial coordinates and the charge of nucleus $I$. 
The wavefunction of electrons obeys Fermi-Dirac statistics thus should be antisymmetric with respect to the interchange of both the spatial coordinates and the spins of any two electrons, namely the following equality of wavefunction should hold: $\psi(\cdots,\bm{\mathrm{x}}_i,\cdots,\bm{\mathrm{x}}_j,\cdots)=-\psi(\cdots,\bm{\mathrm{x}}_j,\cdots,\bm{\mathrm{x}}_i,\cdots)$.\par
Unlike most methods that use variational principle to approach the ground state wavefunction, DMC is a stochastic projection method. A given antisymmetric wavefunction $\psi_T$ can always be represented as a linear combination of a set of eigenfunctions ${\psi_k}$ of the corresponding Hamiltonian operator,
\begin{eqnarray}
    \psi_T(\bm{\mathrm{x}}_1,\cdots,\bm{\mathrm{x}}_n)=&&\sum_{k=0}^{\infty}c_k\psi_k(\bm{\mathrm{x}}_1,\cdots,\bm{\mathrm{x}}_n),\\
    \hat{H}\psi_k(\bm{\mathrm{x}}_1,\cdots,\bm{\mathrm{x}}_n)=&&E_k\psi_k(\bm{\mathrm{x}}_1,\cdots,\bm{\mathrm{x}}_n),\nonumber
\end{eqnarray}
When an imaginary time evolution operator acts on $\psi_T$,
\begin{eqnarray}
    e^{-\tau(\hat{H}-E_T)}\psi_T=\sum_{k=0}^{\infty}c_k e^{-\tau (E_k-E_T)}\psi_k,
\end{eqnarray}
where $E_T$ is the trial energy as an offset, 
there will be a decay coefficient added to each expansion term, and the decay rate is proportional to state energy $E_k$. 
After a long enough imaginary time evolution, $\psi_T$ can reach the ground state $\psi_0$, whereas contributions from all other eigenfuntions vanish. 
If we define a time-dependent wavefunction and look at the imaginary-time Schr\"{o}dinger equation:
\begin{eqnarray}
    \psi(\bm{\mathrm{x}}_1,\cdots,\bm{\mathrm{x}}_n,\tau)=e^{-\tau(\hat{H}-E_T)}\psi_T(\bm{\mathrm{x}}_1,\cdots,\bm{\mathrm{x}}_n),\nonumber\\
    -\partial_\tau\psi(\bm{\mathrm{x}}_1,\cdots,\bm{\mathrm{x}}_n,\tau)=(\hat{H}-E_T)\psi(\bm{\mathrm{x}}_1,\cdots,\bm{\mathrm{x}}_n,\tau).
\end{eqnarray}
Without the potential energy terms, it resembles a standard diffusion equation,
\begin{eqnarray}
    \partial_\tau\psi(\bm{\mathrm{x}}_1,\cdots,\bm{\mathrm{x}}_n,\tau)=\frac{1}{2}\sum_i\nabla_i^2\psi(\bm{\mathrm{x}}_1,\cdots,\bm{\mathrm{x}}_n,\tau).
\end{eqnarray}
The diffusion equation defines the master equation of stochastic processes, hence we can solve the diffusion equation of wavefunction by simulating the stochastic processes \cite{karlin1981second}.
With potential terms, additional processes are required to bind the diffusion equation in simulation (see, e.g., Refs. \cite{umrigar1993diffusion, foulkes_quantum_2001, reynolds1990diffusion} for more details).

\subsection{Trial wavefunction}
In this work we use FermiNet neural network ansatz
as our trial wavefunction. 
Due to the huge number of parameters, it is challenging to converge the training process of FermiNet unless the system is small enough. 
After many tests, we identified a common training pattern of FermiNet, which consists of two stages:
a relatively short sharp-adjustment stage and a lengthy fine-tuning one. We propose to use the FermiNet wavefunction right after the sharp-adjustment stage as the trial wavefunction in DMC, which maximizes the efficiency of the entire simulation protocol. 
In this way we can also achieve more accurate results than a better converged FermiNet model after the lengthy fine-tuning stage.
Comparing to the gain, the cost of performing DMC on the long-trained FermiNet is rather minor in most of the systems tested.

\subsection{DMC implementation}
We have developed a GPU-friendly DMC software in JAX~\cite{jax2018github}, which can be seamlessly integrated with FermiNet~\cite{FermiNet2020github}, developed in the same programming framework.
Our DMC software can also be integrated with other trial wavefunctions implemented in JAX and it has been open sourced in order to accelerate further combination of QMC methods with neural networks.
See Algorithm~\ref{algoDMC} for a brief workflow of one DMC iteration, beyond which various of modifications are implemented, including those proposed by Umrigar et al. to reduce time-step error \cite{umrigar1993diffusion} and by Zen et al. to keep size consistency \cite{zen2016boosting}. 

Random walkers' branching and merging change the total number of walkers, which cause efficiency issue for JAX program and is also not friendly to distributed computing especially when load balancing is involved. We devised a new branching-merging strategy to overcome these issues. Whenever we need to branch certain random walker due to its overly large weight, we also merge two walkers on the same computing node with the smallest weight. No merging is executed if no branching happens. In this way, we keep the number of walkers on each computing node unchanged. We did thorough numerical verification of this strategy and found that the introduced bias is negligible.

The most time-consuming module in our DMC implementation is to calculate the local energy. 
In our optimized program, the computational cost for each local energy estimation is almost same as a VMC inference step of the original FermiNet. Therefore, the total cost depends solely on the number of iterations performed in DMC and VMC.

\begin{algorithm}[htp]
    \caption{Simplified Diffusion Monte Carlo algorithm pseudocode.}\label{algoDMC}
    \KwData{number of walkers $M$, walkers' positions $\bm{\mathrm{r}}$, walkers' weight $\omega$, value and sign of the trial wavefunction  $\Psi_T(\bm{\mathrm{r}})$, $\mathrm{sign}(\Psi_T(\bm{\mathrm{r}}))$, drift velocity $\bm{v}_D(\bm{\mathrm{r}})$, local energy $E_L(\bm{\mathrm{r}})$.}
    \KwOut{accept ratio of trial moves $p_{move}$, trial energy $E_T$, mixed estimated energy $E_D$.}
    \SetKw{KwCompute}{compute}
    \For{$N_{max}^{DMC}$}{
    \For{each walker}{
    $\xi\sim N(0,\tau)$\\
    $\bm{r'}=\bm{\mathrm{r}}+\tau\bm{v}_D(\bm{\mathrm{r}})+\xi$\\
    \KwCompute $\Psi_T(\bm{r'})$, $\mathrm{sign}(\Psi_T(\bm{r'}))$, $\bm{v}_D(\bm{r'})$\\
    $G(\bm{\mathrm{r}}\to\bm{r'})=\exp\left[-(\bm{r'}-\bm{\mathrm{r}}-\tau\bm{v}_D(\bm{\mathrm{r}}))^2/2\tau\right]$\\
    $G(\bm{r'}\to\bm{\mathrm{r}})=\exp\left[-(\bm{\mathrm{r}}-\bm{r'}-\tau\bm{v}_D(\bm{r'}))^2/2\tau\right]$\\
    $p=\mathrm{min}\left\{1,\frac{G(\bm{r'}\to\bm{\mathrm{r}})\Psi_T(\bm{r'})^2}{G(\bm{\mathrm{r}}\to\bm{r'})\Psi_T(\bm{\mathrm{r}})^2}\right\}$\\
    \If{$\mathrm{sign}(\Psi_T(\bm{r'}))\neq\mathrm{sign}(\Psi_T(\bm{\mathrm{r}}))$}{$p=0$}
    $\alpha\sim U(0,1)$\\
    \If{$p\geq\alpha$}{\tcp{Diffuse}
    \KwCompute $E_L(\bm{r'})$\\
    $w\gets w\times\exp[-\tau(E_L(\bm{\mathrm{r}})+E_L(\bm{r'})-2E_T)/2]$\\
    $\bm{\mathrm{r}}\gets\bm{r'}$\\
    $\Psi_T(\bm{\mathrm{r}})\gets\Psi_T(\bm{r'})$\\
    $\mathrm{sign}(\Psi_T(\bm{\mathrm{r}}))\gets\mathrm{sign}(\Psi_T(\bm{r'}))$\\
    $\bm{v}_D(\bm{\mathrm{r}})\gets\bm{v}_D(\bm{r'})$\\
    $E_L(\bm{\mathrm{r}})\gets E_L(\bm{r'})$\\
    }
    }
    \KwCompute $p_{move}$\\
    $E_T=E_T-\log\frac{\sum w}{M}$\\
    $E_D=\frac{\sum w E_L(\bm{\mathrm{r}})}{\sum w}$
    }
\end{algorithm}

\subsection{Energy calculation}
For FermiNet-VMC, we always perform a separate inference simulation for energy estimate,
where we fix all the parameters of FermiNet after training and do a number of Markov Chain Monte Carlo (MCMC) steps to sample batches of random walkers accordingly. We calculate the average local energy for each batch, and use reblock analysis to determine the mean value of the set of averaged energy as well as the standard deviation.
For FermiNet-DMC, we use the mixed estimator of energy \cite{umrigar1993diffusion} and treat the first 10\% of MC steps as the equilibrating phase and only use the steps afterwards for energy production. See Supplementary Table 6-12 for the hyperparameters of all our calculations.
We also use reblock analysis to determine the mean of the averaged energy and its standard error.
In our plots, error bars represent one standard error for energy estimates, unless otherwise specified.

Note that walkers in DMC are more auto-correlated than the ones in VMC inference phase especially when the time step used in DMC is set to be small to avoid bias. Therefore more batches of random walkers are needed to reduce the statistical error to a given level in DMC than in VMC. However in practice, we found that the number of the required extra batches of walkers in DMC is usually much fewer than the number of steps in VMC training phase for full convergence.

\subsection{Nodal structure and wavefunction visualization}\label{sec:method_visualization}
The three dimensional cuts of the full 11D nodal structure of Be  in Fig. 2b-d is plotted according to the rules of Bressanini et al. \cite{bressanini2012implications}.
The four electrons' spherical coordinates are respectively $$
\left\{
\begin{array}{lll}
     r_1\in\left[0.1,\ 2.1\right]\ {\rm a.u.}, & \phi_1=0, & \theta_1\in[-1,\ 1],  \\
     r_2=1.1\ {\rm a.u.}, & \phi_2=0, & \theta_2=\pi/2, \\
     r_3\in\left[0.1,\ 2.1\right]\ {\rm a.u.}, & \phi_3=\pi/2, & \theta_3=\pi/2, \\
     r_4=1.1\ {\rm a.u.}, & \phi_4=3\pi/2, & \theta_4=\pi/2,
\end{array}   
\right.
$$
fixing all the degrees of freedom except $r_1$, $\theta_1$ and $r_3$. The green surfaces in the plots show the nodal surfaces, i.e. the places where the value of wavefunction is zero.

To visualize the nodal surface of benzene, we calculated the wavefunction value on 2-dimensional slices of the 126-dimensional space. 
We first fixed a 126-dimensional electron configuration at the representative position of benzene electronic structure from Liu et al. \cite{Liu2020}, and perturb it slightly for the visualization purpose. 
To construct one slice of the 126-dimensional space, we move a single spin-up electron in a 2-dimensional square with all other 41 electrons fixed. 
Then we apply FermiNet to points on each slice and display the log-scaled magnitude of the evaluated wavefunction value, where the points with small value stand for the nodes on each slice. 
Since the FermiNet output is unnormalized, diagrams for different FermiNet may have drastically different range of displayed value.

\subsection{Divergence measuring nodal surface difference}\label{sec:node_diff}
We define a divergence measuring the difference between two sets $S_1$ and $S_2$ in any metric space as follows
\begin{equation}
    D(S_1, S_2) = E_{Y\sim P_1} d(Y, S_2) \approx \sum_{i}^{K} d(Y_i, S_2)
\end{equation}
where $P_1$ is a probability measure on $S_1$, and $\{Y_i\}_{i=1,...,K}$ are sampled from $P_1$.
The distance $d(Y, S)$ between a single point $Y$ and a set $S$ is defined as the smallest distance between $Y$ and any point in $S$, namely
\begin{equation}
    d(Y, S) = \min_{Z\in S} d(Y, Z)
\end{equation}

For a nodal surface S corresponding to an unnormalized wavefunction $\Psi$, we would like to define a measure on S such that a small area on S is assigned  larger weight if its neighbourhood has larger $\Psi^2$ value, namely larger probability to be visited by walkers in DMC.
Therefore, we consider a neighbourhood 
\begin{equation}
    S_{\epsilon} = \{x | d(x, S) < \epsilon\}
\end{equation}
\\
around S and a mapping 
\begin{equation*}
\phi: S_\epsilon \rightarrow S,
\end{equation*}
then "push forward" the probability density $m_{\Psi^2}$ (corresponding to $\Psi^2$) from $S_\epsilon$ to S via $\phi$, namely
\begin{equation}\label{eq:node_measure}
\phi\circ m_{\Psi^2} (n) \coloneqq m_{\Psi^2} (\phi^{-1}(s)) = \frac{\int_{\phi^{-1}(s)}\Psi^2}{\int_{S_\epsilon}\Psi^2}, \quad \forall \text{ set }s\subset S
\end{equation} 
Intuitively, for any point $y$ in $S_\epsilon$ we may simply choose $\phi(y)$ to be the point on $N$ that is closest to $y$.

However, it's quite difficult to determine both $S_\epsilon$ and $\phi$ mentioned above algorithmically, and thus, in practice, we use some approximate alternatives that are much easier to compute.
See Supplementary Note 15 for the algorithmic detail. 

\subsection*{Data Availability}
All data supporting the findings of this study are provided in Supplementary Information.

\subsection*{Code Availability}
We have released our DMC software at \url{https://github.com/bytedance/jaqmc}. 

\bibliography{ref}

\subsection*{Acknowledgments}

\begingroup
\footnotesize
This work is directed and supported by Hang Li and ByteDance Research. We thank Yubing Qian for providing preliminary settings for water dimer simulations. We thank Weinan E, Xiang Li, Kai Zheng, Ke Liao and Yu Liu for fruitful discussions. We thank Mike Entwistle and James Spencer for sharing data and results. We thank Michel Caffarel for allowing us to use an adapted version of the figure on his website. We thank Shaochen Shi, Xiaoying Jia, Xin Liu and Chenlin Chai for engineering improvement on our DMC software. We thank the rest of ByteDance Research team for inspiration and encouragement. 
J.C. is supported by the National Key R\&D Program of China under Grant No. 2021YFA1400500, the National Natural Science Foundation of China under Grant No. 92165101 and No. 11974024, and the Strategic Priority Research Program of Chinese Academy of Sciences under Grant No. XDB33000000.

\subsection*{Author contributions}

\begingroup
\footnotesize
W.R. and J.C. conceived the study; W.R. implemented the main code with important contributions from W.F.; W.R. and W.F. performed simulations, data analyses, and figure designing. X.W. performed CCSD(T) related calculation. J.C. supervised the project. W.R., W.F., X.W., and J.C. wrote the paper.
\endgroup

\subsection*{Competing interests}
The authors declare no competing interests.

\end{document}


\title{Supplementary Information: Towards the ground state of molecules via diffusion Monte Carlo on neural networks}


\author[1]{Weiluo Ren}
\author[1,2]{Weizhong Fu}
\author[1]{Xiaojie Wu}
\author[2,3]{Ji Chen}
\affil[1]{ByteDance Research, Zhonghang Plaza, No. 43,  North 3rd Ring West Road, Haidian District, Beijing, People’s Republic of China}
\affil[2]{School of Physics, Peking University, Beijing 100871, People’s Republic of China}
\affil[3]{Interdisciplinary Institute of Light-Element Quantum Materials, Frontiers
Science Center for Nano-Optoelectronics, Peking University, Beijing 100871, People’s Republic of China}

\maketitle

\section{Additional software detail}

In the method section, we have briefly described the DMC software developed for this work, including key methodological details. 
Here we add a few more points, and a complete description of our software will be published elsewhere. 

Our implementation of DMC is scaling-out friendly. 
We have tested our software on a GPU cluster with 16 computing nodes and 128 Nvidia V100 GPU cards, and obtained near-linear speed-up.
Our software saves checkpoint for DMC periodically and uploads to a remote storage cluster so that it can be resilient to occasional preemption due to cluster scheduling and device failure, and it is able to continue the process with the most recent checkpoint downloaded when a rescheduling is triggered. 

In order to support multi-node computation efficiently, we have to minimize the inter-nodes communication, listed below:
\begin{itemize}
    \item We calculate the total energy with local energy and weights for all the walkers across all the nodes.
    \item We also calculate several metrics with across-node aggregation, such as acceptance rate, number of outliers and etc.
    \item We synchronize the action of uploading and downloading checkpoints.
\end{itemize}
We have not yet supported across-node load balancing in terms of walkers, which could cause large communication overhead. 
As a matter of fact, we have devised a branching-merging algorithm which fixes the walker size per computing node so that we do not need to apply across-node walker load balancing at all. 
Every time when we do branching at the presence of a large-weight walker, we simultaneously find two walkers on the same node but with the smallest weight and merge them. Note that if the trial wavefunction has good quality and we have a decent number of walkers on each node, then such an action does not introduce much bias, confirmed in our calculations and shown in Supplementary Note 3.
We only use the fixed-size branching scheme for relatively large systems, such as cyclobutadience, benzene molecule and benzene dimers, in which cases efficiency is crucial due to the computation power limit and the fixed-node error dominates the extra bias introduced by this branching scheme. 

In order to revert the effect of changing trial energy, the energy produced from each DMC step should be corrected accordingly before averaged.
However, as pointed in \cite{needs2019user}, when the batch size is large enough, the effect of such correction is negligible, which is consistent with our experience. Furthermore, it is easier to do reblocking analysis \cite{flyvbjerg1989error} to assess statistical error without such correction. 
Therefore, for all the results reported here, we do not include such energy corrections in our computation.

We have also implemented the electron-by-electron scheme suggested by CASINO \cite{needs2020variational}, where each walker moves electrons one by one rather than all at once. 
Our tests show that the electron-by-electron scheme leads to a significant increase in the acceptance ratio of walkers with large time-steps, but we do not see much improvement on the time-step error or auto-correlation period.
Besides, our electron-by-electron implementation sequentially goes over all the electrons in one configuration, leading to a decline in computing efficiency when dealing with large systems. 
So this option is also switched off by default.

\section{Finite time-step error}\label{sec:finite_time_error}
We have implemented the ZSGMA approach proposed by Zen et al. to reduce the finite time-step error \cite{zen2016boosting}.
As tested in various systems in literature our default setting $10^{-3}$ is usually small enough to reach convergence.
Here we show a test on the cyclobutadience  with equilibrium structure as an example. The results are displayed in Supplementary Fig. \ref{fig:dmc_si_finite_timestep_error}, from which we can conclude that the finite time-step error for our default setting $10^{-3}$ is well below 1 mHa, comparable to the statistical error. 
Based on this observation as well as the efficiency concern, we use time-step $10^{-3}$ by default and use it in all of our FermiNet-DMC calculations.

To obtain comparable statistical error of the reported energy, we use different number of DMC steps for each time-step to compensate different level of auto-correlation in the energy time series. 
For time-step $10^{-3}$, we run DMC for $10^5$ steps.
For large time-steps like 0.01 or 0.03, we only run DMC for $10^4$ steps while for small time-step like $10^{-4}$, we run DMC for $10^6$ steps.

\begin{figure}
  \centering
  \includegraphics[width=144mm]{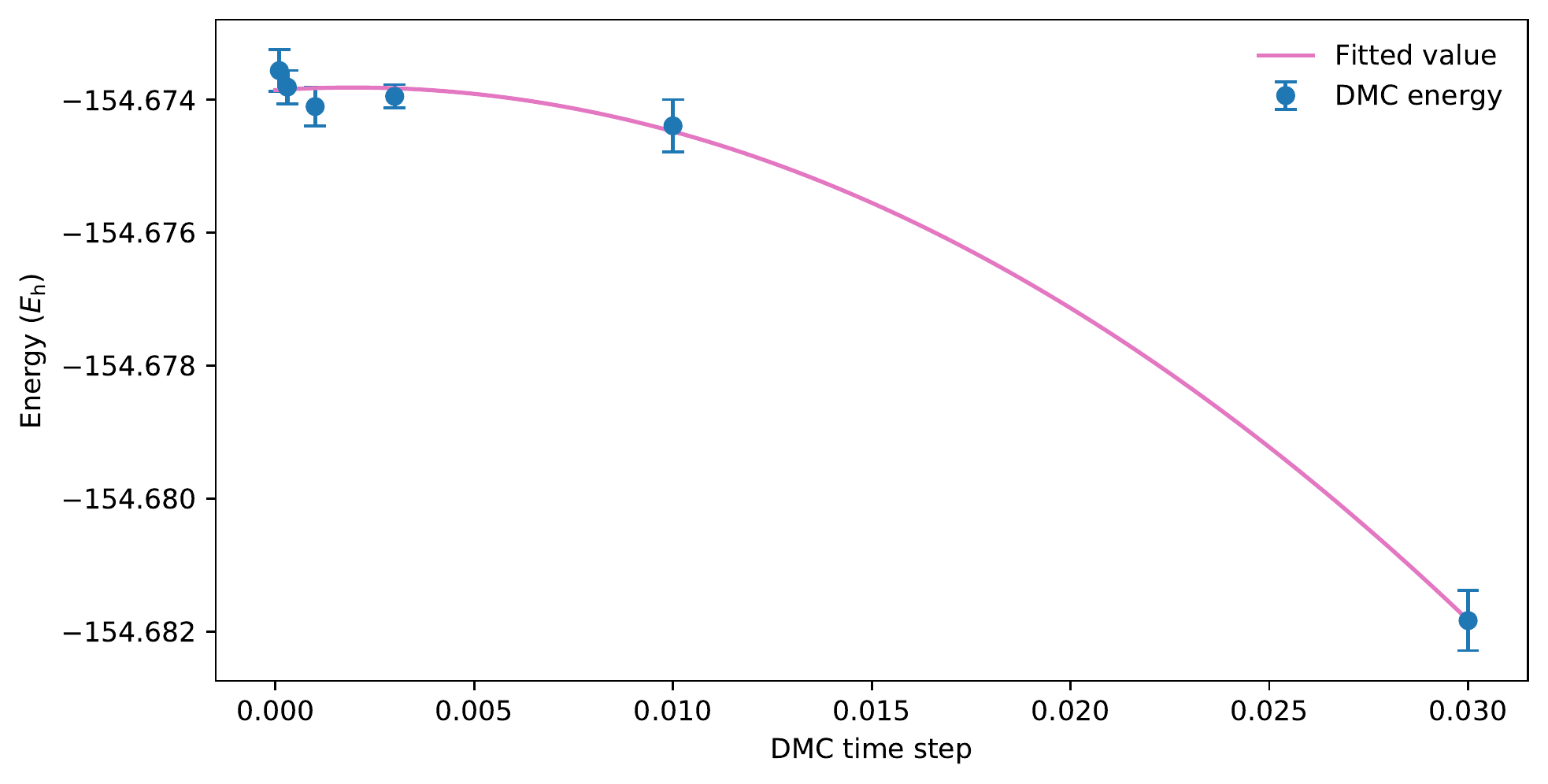}
  \caption{\label{fig:dmc_si_finite_timestep_error}Time-step dependence of the total energy of cyclobutadience with equilibrium structure. The pink curve is generated by doing a quadratic fitting on the collected DMC energy data points.}
\end{figure}

\section{DMC options}

To investigate the effect of different hyperparameters of our DMC software, especially the bias introduced in different settings, we compare the DMC energy calculated with the settings listed in Supplementary Table \ref{tab:dmc_si_errorbar_settings}, using cyclobutadience with equilibrium structure as an example. 
The trial wavefunction is a FermiNet trained with $10^5$ steps. 
See Note \ref{sec:cb} for more training details.
The result is displayed in Supplementary Fig. \ref{fig:dmc_si_errorbar}. 
Taking the statistical error into account, all settings agree with each other well except the most inaccurate one with time-step $10^{-2}$.

\begin{table*}[htb]
\centering
\caption{\label{tab:dmc_si_errorbar_settings} We test our DMC software with different hyperparameters, listed as follows. ``time step'' is the DMC time-step. ``equilibrium phase length'' is the length of phase in the beginning of the whole DMC process during which we do not collect statistics for report purpose. ``total number of steps'' is the number of steps in the whole DMC process. ``energy window size'' is the length of the rolling window for trial-energy update. ``fix size'' indicates whether we fix the number of DMC walkers on each computing node or not. ``energy cutoff'' is the parameter $\alpha$ in the cutoff scheme proposed in \cite{zen2016boosting} and indicates the level of energy clipping.}
\begin{tabular}{lcccccc}
\toprule
     \thead{Identifier} & \thead{time step}& \thead{equilibrium phase length} & \thead{total number of steps}& \thead{energy window size}  & \thead{fix size} & \thead{energy cutoff} \\
\midrule
     time step $10^{-2}$ & $10^{-2}$ & $2\times 10^3$ &$10^4$& $10^3$ & True & 2\\
     time step $10^{-3}$ (default) &$10^{-3}$ & $2\times 10^4$ &$10^{5}$  & $10^4$ & True & 2  \\
     time step $10^{-4}$ & $10^{-4}$& $2\times 10^5$ & $10^6$ & $10^5$ & True & 2 \\
     window $5 \times 10^4$ & $10^{-3}$ & $10^5$ & $2\times10^5$  & $5 \times 10^4$ & True & 2 \\
     vary size & $10^{-3}$ & $2\times 10^4$ & $10^5$ & $10^4$ & False & 2 \\
     energy cutoff 4 & $10^{-3}$ & $2\times 10^4$ & $10^5$ & $10^4$ & True & 4 \\
\bottomrule
\end{tabular}
\end{table*}

\begin{figure}
  \centering
  \includegraphics[width=144mm]{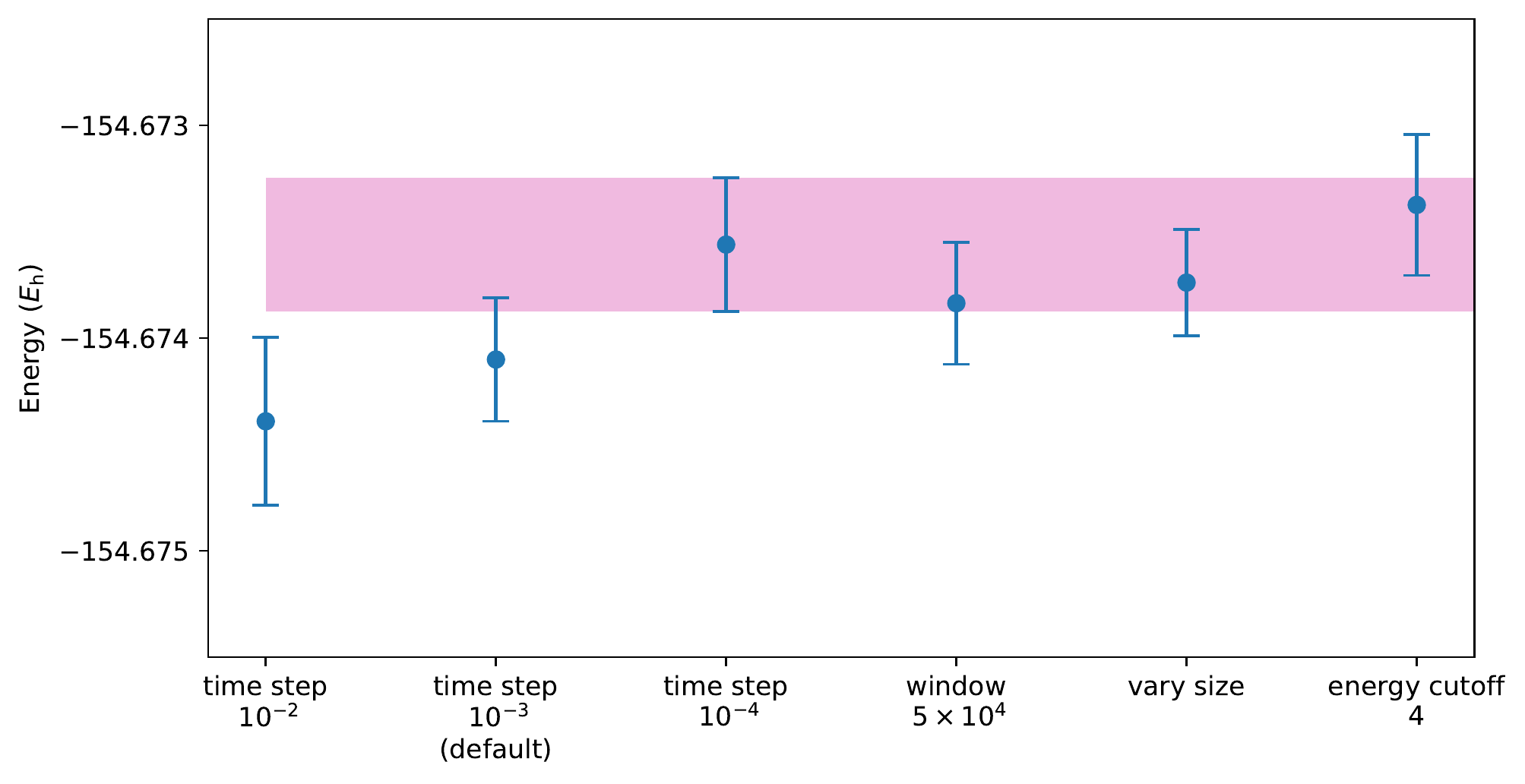}
  \caption{\label{fig:dmc_si_errorbar}DMC energy and statistical error for different settings listed in Supplementary Table \ref{tab:dmc_si_errorbar_settings}. The pink patch indicates the one-standard-error interval of energy in the most accurate setting with time step $10^{-4}$. All the error bars are overlapped with this interval except the most inaccurate setting with time-step $10^{-2}$.}
\end{figure}

\section{Efficiency}
In this section, we compare the efficiency between FermiNet-VMC \cite{ferminet,spencer_better_2020,FermiNet2020github} and FermiNet-DMC. 
The evaluation of FermiNet's local energy dominates the whole FermiNet-DMC process in terms of computation time. 
In other words, for one step, the difference between runtime of FermiNet-DMC and the inference phase of FermiNet-VMC is negligible. 

However, DMC usually requires more steps to reduce the statistical error to a given level compared to VMC inference phase, because DMC has to use small time-step to avoid finite time-step error while VMC can use large time-step in Markov chain Monte Carlo (MCMC) to obtain less correlated samples. 
To cut down the statistical error more effectively in DMC, we can increase the batch size so that each batch contains more independent walkers.
In practice, we start FermiNet-DMC procedure from the FermiNet-VMC walkers. 
To increase the DMC batch size, we duplicate the VMC walkers and then do MCMC for a period of time to get a larger batch of uncorrelated walkers, used as the starting walkers in DMC. 
Even if we do not enlarge the batch size, we still do a number of MCMC steps using fixed FermiNet before DMC so that the walkers is distributed closer to the square of FermiNet's wavefunction, which serve better starting point for DMC.
Note that doing MCMC with FermiNet is much faster than doing DMC since no local energy calculation is involved. In our experiments, we use larger DMC batch size only when handling benzene dimers, in which case DMC uses batch size 16384 and VMC uses batch size 4096. 

Overall FermiNet-DMC is still a much more efficient solution than FermiNet-VMC, because it can drastically cut down the length of the training phase of VMC which is significantly longer than the DMC itself. As a matter of fact, in the calculations that we performed, FermiNet-DMC can reach the same or even better accuracy level with a FermiNet trained with only one tenth or even a smaller fraction of total number of steps compare to the fully converged FermiNet-VMC.

\section{Details on large systems}\label{sec:large_system}
When dealing with relatively large systems containing dozens if not hundreds of electrons, chances are that we can not afford a large enough network to well represent the ground state wavefunction. 
Several issues emerge in this regime when training FermiNet. 
Firstly, the under-fit FermiNet may hit outliers hurting the convergence process. 
What's worse, the training process may hit NAN issue occasionally. 
We removed the outliers in terms of local energy when calculating the gradient in the FermiNet's training process.
This helps to stabilize the calculation of the average and standard deviation of local energies in a batch, which benefits the gradient clipping process. 
Without such outlier-removal operation, at the presence of one or more outliers, the gradient clipping is not effective since we are not able to determine the clipping level correctly.

Secondly, the determinant calculation of large-dimension matrix is numerically troublesome.
Potentially, there could be overflow and underflow issues especially when using single precision.
Besides that, there is usually a dominating determinant among multiple determinants, which makes the value of other determinants effectively zero due to float number rounding. 
Therefore, we only use a single determinant when dealing with large systems so that the process becomes more efficient and stable.
Admittedly, this limits the network expressiveness, and further improvements are left for future work.

Thirdly, the under-fitted decaying behavior of electron density at a distance may cause serious issues, introducing outliers in the calculation of total energy. 
For instance, there could be walkers occasionally diffusing away from atoms and reaching areas where the neural network is not well trained, which causes trouble for the calculation of total energy and its gradient. 
We use isotropic envelope \cite{spencer_better_2020} to maintain the exponentially decaying behavior of the wavefunction when the electron is away from all the atoms, which also helps to improve the efficiency of both FermiNet-VMC and FermiNet-DMC.

\section{Common Hyperparameters}
All of our calculations follow the same procedure. 
First we train FermiNet with VMC\cite{ferminet,spencer_better_2020,FermiNet2020github}, then we use the trained FermiNet as trial wavefunction in DMC. 
To obtain reliable energy from VMC, we run a separate inference process with the trained FermiNet. 
There are some hyperparameters shared by all the calculations in this work, as listed in Supplementary Table \ref{tab:common_hyper}.

\begin{table}[htb]
\caption{Common hyperparameters}\label{tab:common_hyper}
\centering
\begin{tabular}{lclc}
\toprule
Hyperparameter & Value & Hyperparameter & Value \\
\midrule
VMC optimizer & KFAC \cite{martens2015optimizing,kfac-jax2022github} & Batch size & 4096 \\
Precision & Float32 & Full determinant & True \\
DMC time-step & $10^{-3}$ &&\\ 
\bottomrule
\end{tabular}
\end{table}

\section{Atoms}
A rather small neural network is used in calculations of single atoms (see Fig. 1e of main text), and all the training processes have well converged. 
The corresponding hyperparameters are listed in Supplementary Table \ref{tab:atom_hyper}.

\begin{table}[htb]
\caption{Hyperparameters for single atoms calculations in Fig. 2e. 
}\label{tab:atom_hyper}
\centering
\begin{tabular}{lclc}
\toprule
Hyperparameter & Value & Hyperparameter & Value  \\
\midrule
Dimension of one electron layer  & 32 & Dimension of two electron layer & 4 \\
Number of layers  & 2 & Number of determinants & 1 \\
Envelope type & full & VMC learning rate & $5\times10^{-4}$ \\
Number of training steps & $\geq5\times10^5$ & Number of inference steps & $10^5$\\
Number of DMC steps & $2\times10^5$ & MCMC steps between each iterations & 100\\
Outlier removal & false & Fixed-size branching & false \\
\bottomrule
\end{tabular}
\end{table}

\section{Nitrogen Molecule}
Note that our FermiNet-VMC result shown in Fig. 3a of main text is better than the reported FermiNet result in \cite{ferminet}.
The main difference is that our calculation uses the full determinant mode, which boosts the accuracy of FermiNet-VMC, as pointed out in \cite{spencer_better_2020,lin2021explicitly}.
\begin{table}[htb]
\caption{Hyperparameters for nitrogen dissuciation curve calculations in Fig. 3a of main text. 
}\label{tab:n2_hyper}
\centering
\begin{tabular}{lclc}
\toprule
Hyperparameter & Value & Hyperparameter & Value  \\
\midrule
Dimension of one electron layer  & 256 & Dimension of two electron layer & 32 \\
Number of layers  & 4 & Number of determinants & 16 \\
Envelope type & full & VMC learning rate & $10^{-4}$ \\
Number of training steps & $2\times10^5$ & Number of inference steps & $10^5$\\
Number of DMC steps & $2\times10^5$ & MCMC steps between each iterations & 10\\
Outlier removal & false &Fixed-size branching & false\\
\bottomrule
\end{tabular}
\end{table}

\section{Cyclobutadiene}\label{sec:cb}
We use the default network structure in the open-source FermiNet repository for cyclobutadiene, and train two neural networks for the equilibrium and transition state configurations separately.
The training process was converged after $5 \times 10^5$ steps, as shown in Supplementary Fig. \ref{fig:cb_si}. 
We use FermiNet trained with $10^5$ and $5 \times 10^5$ steps, and run FermiNet-DMC and FermiNet-VMC inference processes accordingly. 
The results are listed in Supplementary Table \ref{tab:cb_energy}, where we also include the results from \cite{spencer_better_2020} for comparison.

\begin{table}[htb]
\caption{Hyperparameters for cyclobutadiene related calculations}\label{tab:cb_hyper}
\centering
\begin{tabular}{lclc}
\toprule
Hyperparameter & Value & Hyperparameter & Value  \\
\midrule
Dimension of one electron layer  & 256 & Dimension of two electron layer & 32 \\
Number of layers  & 4 & Number of determinants & 16 \\
Envelope type & full & VMC learning rate & $10^{-4}$ \\
Number of training steps & $5\times10^5$ & Number of inference steps & $10^5$\\
Number of DMC steps & $10^5$ & MCMC steps between each iterations & 10\\
Outlier removal & false &Fixed-size branching & true\\

\bottomrule
\end{tabular}
\end{table}

\begin{table}[htb]
\centering
\caption{Cyclobutadiene ground state energy in Hartree. ``VMC'' and ``DMC'' indicates our FermiNet-VMC and FermiNet-DMC results respectively. For FermiNet-DMC, we include results corresponding to FermiNet trained with $10^5$ and $5\times10^5$ steps. ``${\rm VMC}^*$'' indicates results from \cite{spencer_better_2020}.}\label{tab:cb_energy}
{\renewcommand{\arraystretch}{1.2}%
\begin{tabular}{@{\extracolsep{\fill}}lccccc@{\extracolsep{\fill}}}
\toprule
\multirow{2}{*}{Configuration}&\multicolumn{2}{@{}c@{}}{VMC}&\multicolumn{2}{@{}c@{}}{DMC} & \multirow{2}{*}{${\rm VMC}^*$}\\
\cmidrule{2-3}\cmidrule{4-5}%
 & $10^5$ step &$5\times10^5$ step &  $10^5$ step &  $5\times10^5$ step & \\
\midrule
Equilibrium & -154.6577(1) & -154.6655(1) & -154.6740(3) &-154.6770(2) & -154.6719(1)\\ 
Transition & -154.6430(1) & -154.6505(1) & -154.6585(3) & -154.6611(2) & -154.6554(1)\\
\bottomrule

\end{tabular}
}
\end{table}

\begin{figure}
  \centering
  \includegraphics[width=144mm]{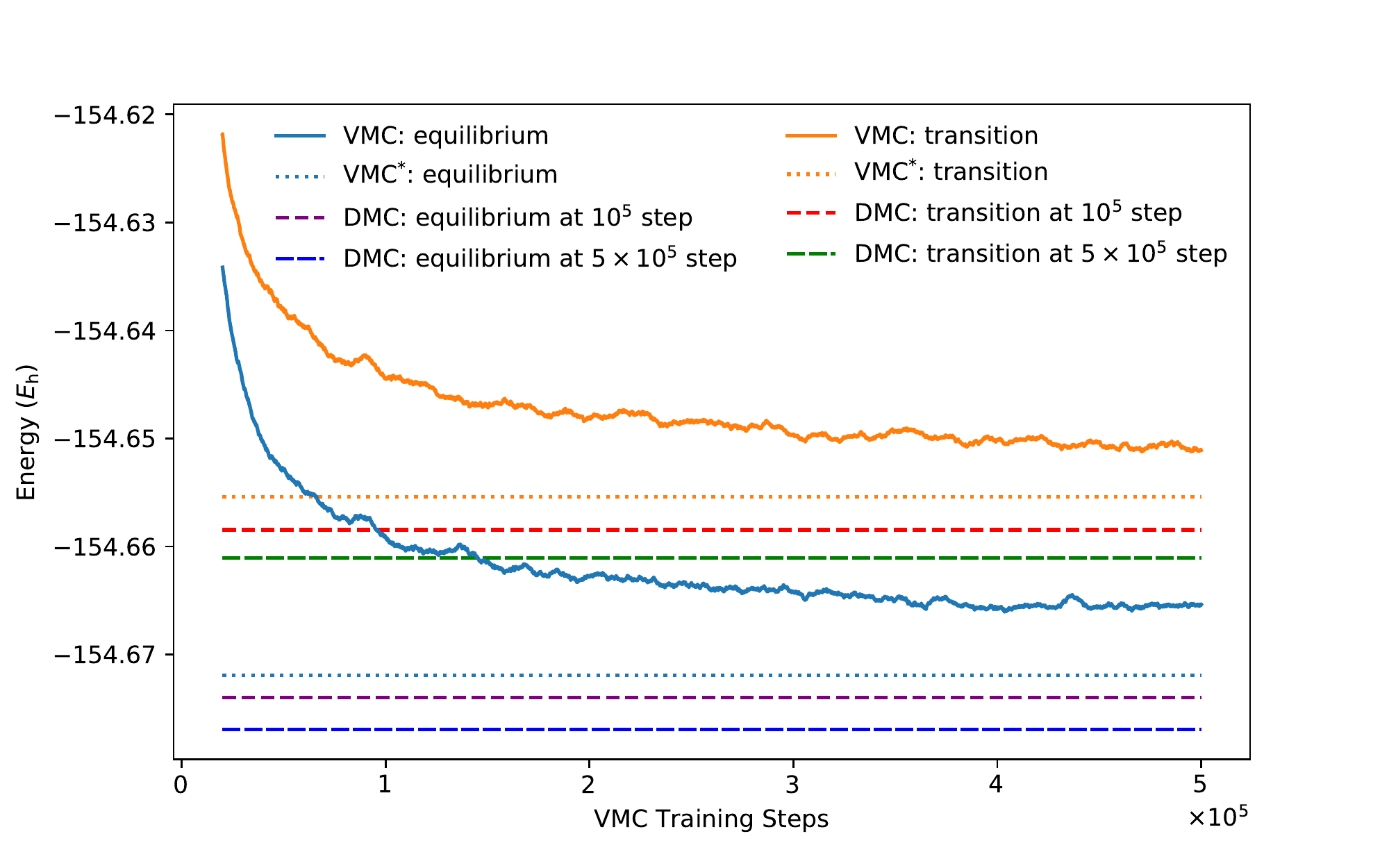}
  \caption{\label{fig:cb_si} The FermiNet-VMC training curve and FermiNet-DMC energy for cyclobutadience equilibrium and transition configurations. ``VMC'' and ``DMC'' indicates our FermiNet-VMC and FermiNet-DMC results respectively. For FermiNet-DMC, we show results corresponding to FermiNet trained with $10^5$ and $5\times10^5$ steps. ``${\rm VMC}^*$'' indicates results from \cite{spencer_better_2020}. }
\end{figure}

\section{Water Dimer}
The 10 structures of water dimer and their point-group symmetries are shown in Supplementary Fig. \ref{fig:water_si}. 
The energy curves of FermiNet-VMC and corresponding FermiNet-DMC energy are shown in Supplementary Fig. \ref{fig:water_dimer_energy}, where all the VMC training curves show a clear ``elbow'' pattern and transition to the slow-converging phase around $10^5$ step.
The FermiNet-DMC energy at $10^5$ step is already lower than the corresponding FermiNet-VMC energy at $3 \times 10^5$ step for all structures.
Actually, for most structures the VMC energy curve is still not converged at $3\times10^5$ step.
In principle, convergence in VMC training stage is necessary to investigate the energy difference.
All the relative energy results are shown here in Supplementary Fig. \ref{fig:water_dimer_relative_energy}.
In Fig. 3d of main text we see that the mean absolute deviation of the relative energy results of FermiNet-DMC are much better than VMC's, especially at $10^5$ step when the networks are undertrained.
This is reasonable as the nodal surface, which contains the zero order information of wave function, naturally requires fewer training steps to converge.
In conclusion, to calculate the relative energy, DMC is a much more efficient and reliable choice compared with VMC.

\begin{table}[htb]
\caption{Hyperparameters for water dimer calculations in Fig. 3d of the main text. 
}\label{tab:water_hyper}
\centering
\begin{tabular}{lclc}
\toprule
Hyperparameter & Value & Hyperparameter & Value  \\
\midrule
Dimension of one electron layer  & 256 & Dimension of two electron layer & 32 \\
Number of layers  & 4 & Number of determinants & 16 \\
Envelope type & full & VMC learning rate & $10^{-4}$ \\
Number of training steps & $3\times10^5$ & Number of inference steps & $10^5$\\
Number of DMC steps & $2\times10^5$ & MCMC steps between each iterations & 100\\
Outlier removal & false & Fixed-size branching & false\\
\bottomrule
\end{tabular}
\end{table}

\begin{figure}[htb]
  \centering
  \includegraphics[width=15cm]{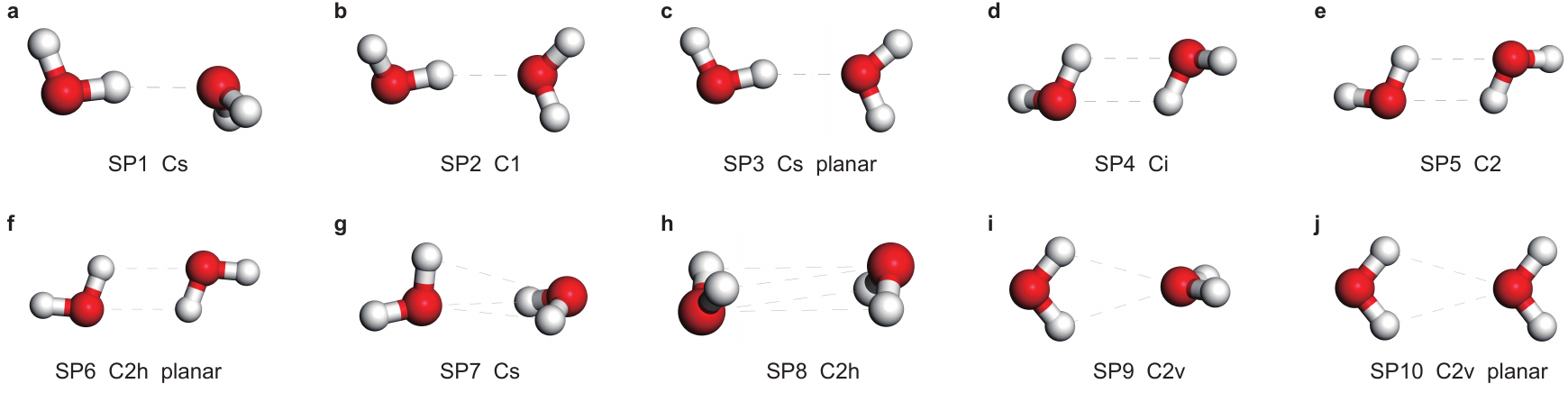}
  \caption{\label{fig:water_si}Geometries of the SP$n$ ($n=1,2,...,10$) structures of water dimer, with their point-group symmetry labels.}
\end{figure}
\begin{figure}[htb]
  \centering
  \includegraphics{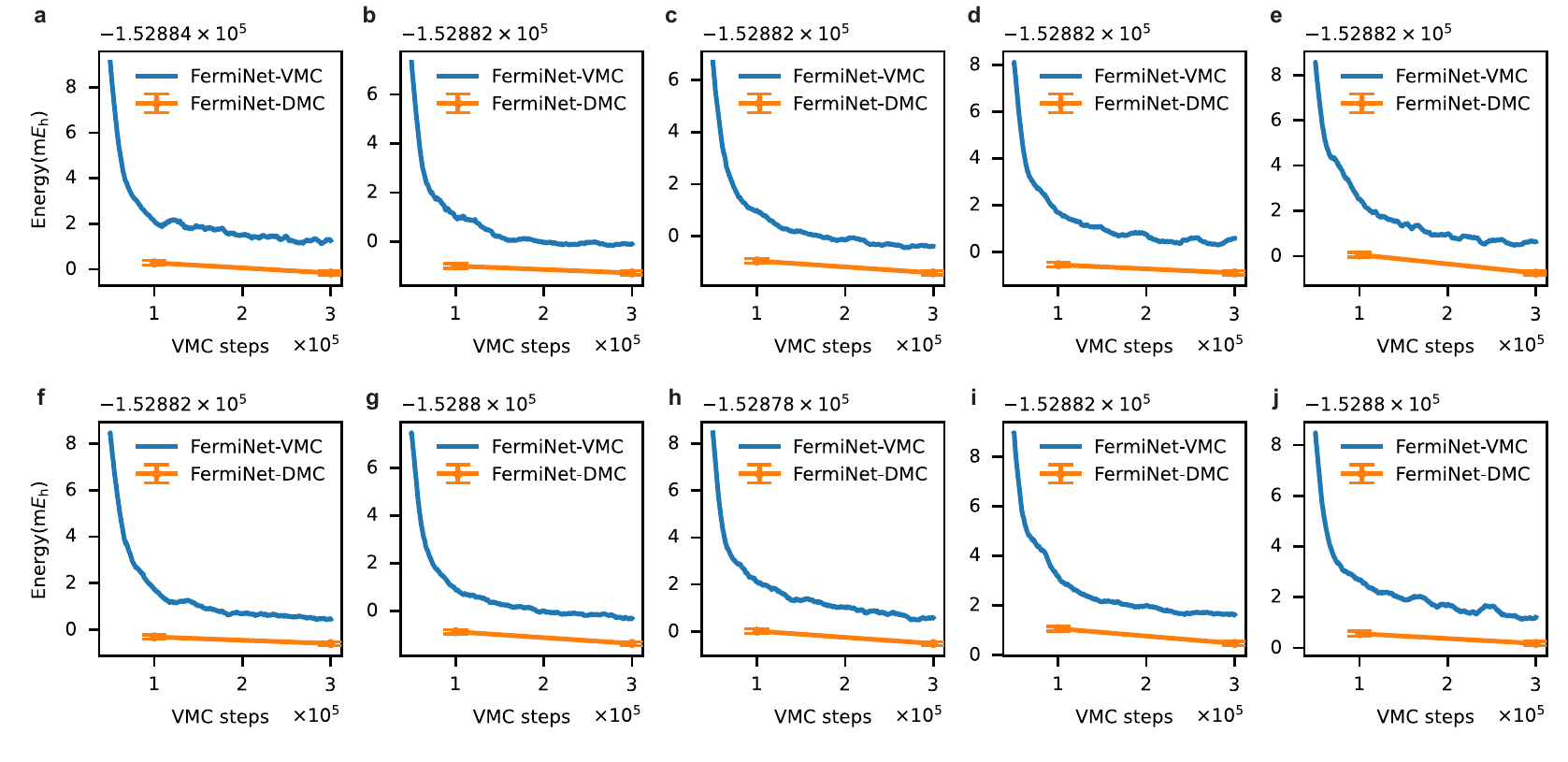}
  \caption{\label{fig:water_dimer_energy}Total energy results of the SP$n$ ($n=1,2,...,10$) structures of water dimer.}
\end{figure}
\begin{figure}[htb]
  \centering
  \includegraphics[width=174mm]{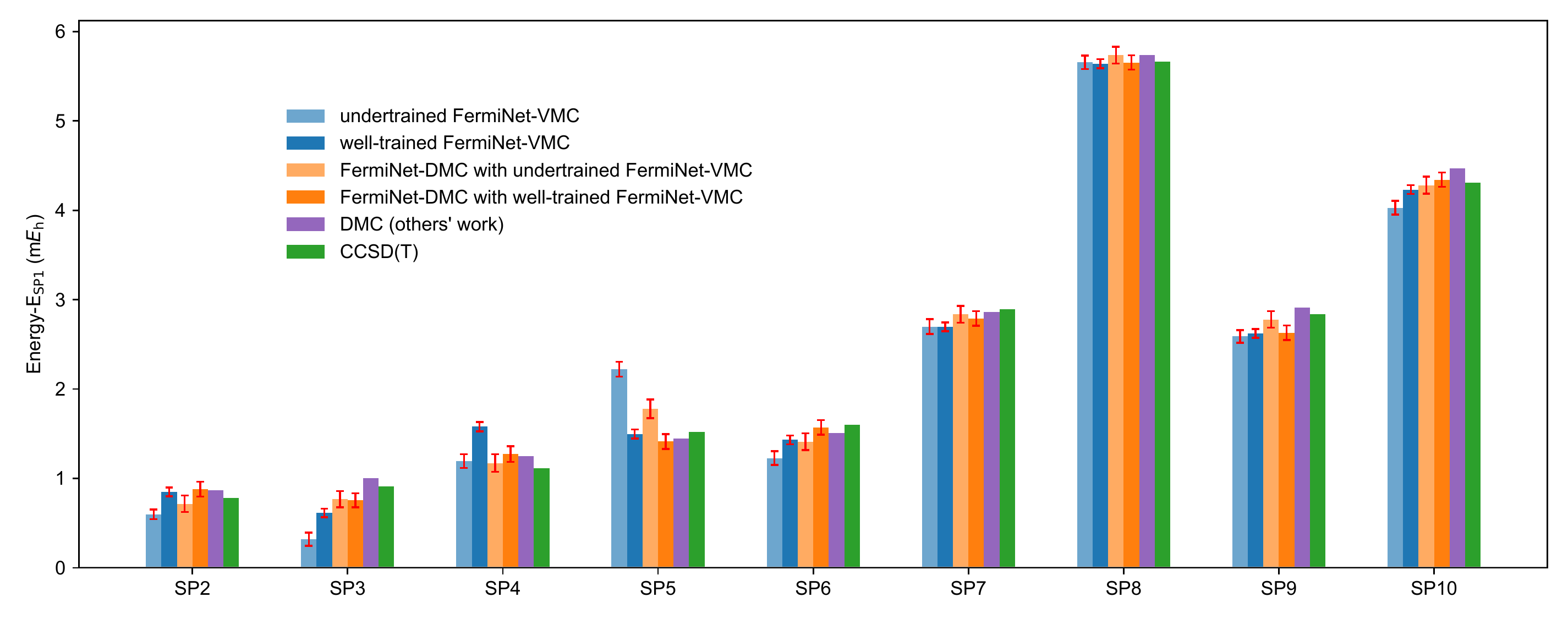}
  \caption{\label{fig:water_dimer_relative_energy}Relative energy results of the SP$n$ ($n=2,3,...,10$) structures of water dimer to the SP1 structure.}
\end{figure}

\section{Benzene}\label{sec:benzene}
We trained a 3-layer and a 4-layer FermiNet for the benzene molecule. The common hyperparameters for both networks are listed in Supplementary Table \ref{tab:benzene_common_hyper}. The hyperparameters indicating the network quality are described in Supplementary Table \ref{tab:benzene_l3_hyper} and \ref{tab:benzene_l4_hyper}, respectively. The energy calculated are listed in Supplementary Table \ref{tab:benzene_energy}.

\begin{table}[htb]
\caption{Common hyperparameters for benzene related calculations}\label{tab:benzene_common_hyper}
\centering
\begin{tabular}{lclc}
\toprule
Hyperparameter & Value & Hyperparameter & Value  \\
\midrule
Number of determinants & 1 & Envelope type & isotropic \\
Number of VMC inference steps & $10^5$ & Number of DMC steps & $2\times10^5$ \\
Learning rate & 0.1 & MCMC steps between each iterations & 100  \\
Outlier removal & True &Fixed-size branching & true \\
\bottomrule
\end{tabular}
\end{table}

\begin{table}[htb]
\caption{Hyperparameters for 3-layer FermiNet for the benzene molecule}\label{tab:benzene_l3_hyper}
\centering
\begin{tabular}{lclc}
\toprule
Hyperparameter & Value & Hyperparameter & Value  \\
\midrule
Dimension of one electron layer  & 128 & Dimension of two electron layer & 8 \\
Number of layers  & 3 & Number of VMC training steps & $2\times 10^6$ \\
\bottomrule
\end{tabular}
\end{table}

\begin{table}[htb]
\caption{Hyperparameters for 4-layer FermiNet for the benenze moledule}\label{tab:benzene_l4_hyper}
\centering
\begin{tabular}{lclc}
\toprule
Hyperparameter & Value & Hyperparameter & Value  \\
\midrule
Dimension of one electron layer  & 256 & Dimension of two electron layer & 32 \\
Number of layers  & 4 & Number of VMC training steps & $10^6$\\
\bottomrule
\end{tabular}
\end{table}

\begin{table}[htb]
\caption{Calculated benzene energy with FermiNet based VMC and DMC in Hartree. The 3-layer neural network is trained with $2\times 10^6$ steps and the 4-layer one is trained with $10^6$ steps. The DMC energy has well converged.}\label{tab:benzene_energy}
\centering
\begin{tabular}{ccc}
\toprule
Network Structure &  FermiNet-VMC & FermiNet-DMC\\
\midrule
3-layer & -232.2143(1) & -232.2330(3) \\
4-layer &-232.2233(1) & -232.2370(3)\\
\bottomrule
\end{tabular}
\end{table}

In Supplementary Fig. \ref{fig:benzene_nodes_CH}, 
we provide more visualization of the log scaled magnitude of wavefunction on slices of the electronic configuration space 
as supplement to the main text Fig. 4.
As described in main text Section II F, the slices are generated by moving a single spin-up electron while keeping all others fixed in their representative positions.
In Supplementary Fig. \ref{fig:benzene_nodes_CH} , the moving electrons are corresponding to the C-H bonds in a benzene. Note that there is no symmetry built in the neural network, while we can still see clear patterns across different subplots in Supplementary Fig. \ref{fig:benzene_nodes_CH}, indicating that FermiNet successfully learned the symmetric property of this system. 
Moreover, the node structure near the light area, namely the one with larger wavefunction magnitude and larger probability to be sampled in the VMC process, shares higher level of similarity. 
Intuitively, this part of nodes plays more important roles in our fixed-node DMC compared to the rest of nodes which are less likely to be visited in the DMC process, further explaining why the 3-layer and 4-layer FermiNet-DMC results are so close.

\begin{figure}[t]
  \centering
  \includegraphics[width=174mm]{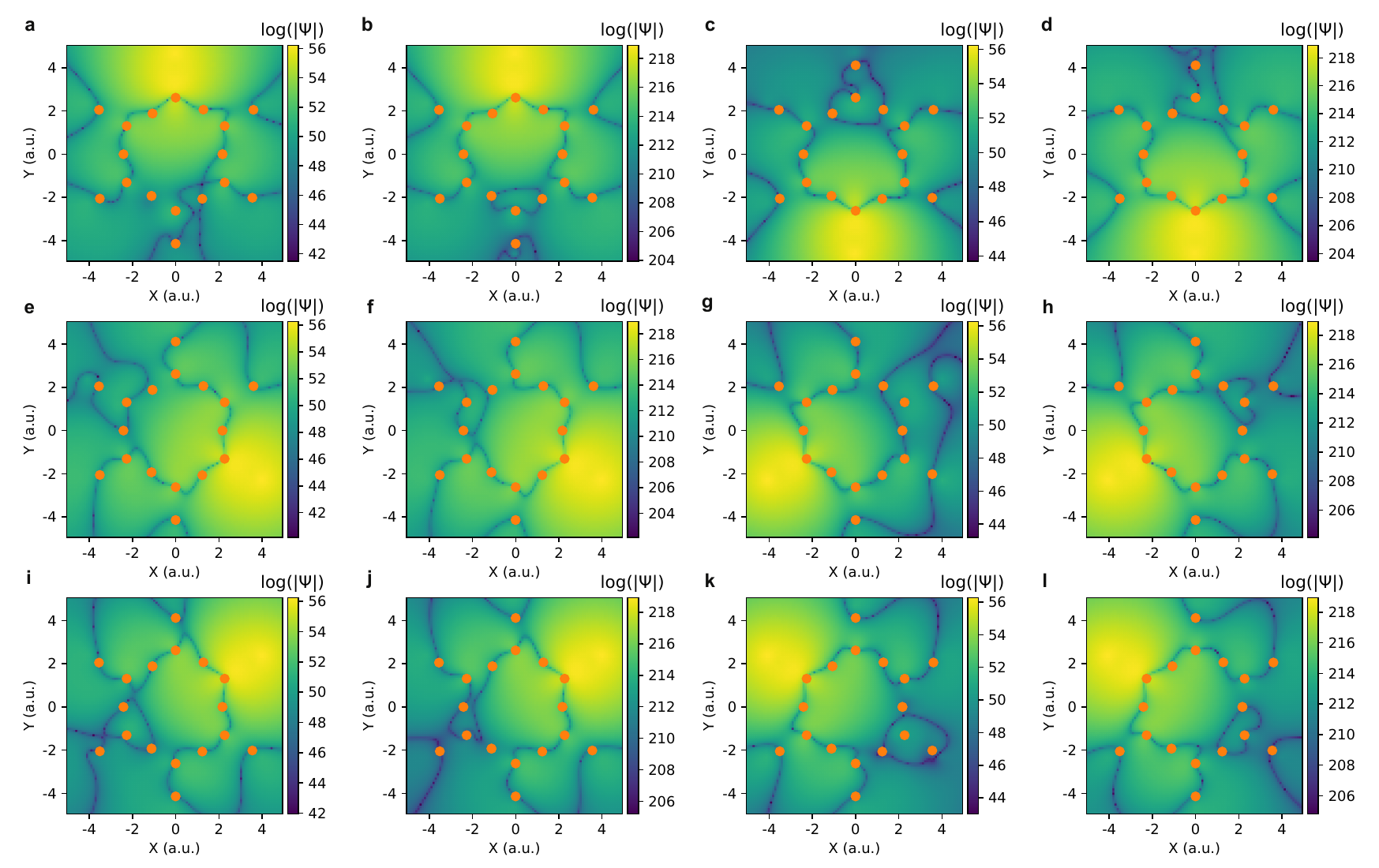}
  \caption{\label{fig:benzene_nodes_CH} The 2-dimensional slices of log scaled magnitude of wavefunctions for benzene. The first and third column of figures are for a 4-layer FermiNet and the rest is for a 3-layer FermiNet. Each figure corresponds to a spin-up electron on a C-H bond, whose position is indicated by the lightest area in the figure. The dark pixels correspond to nodes.}
\end{figure}

\section{Benzene dimer}\label{sec:benzene-dimer}

We use the same 3-layer neural network for the benzene molecule to handle benzene dimers.
%
The network for a benzene dimer is significantly more difficult to train than the benzene molecule. 
One of the tricky hyperparameters is 
the learning rate. 
Small learning rates lead to extremely slow convergence as well as trapping in local minimum, which was also mentioned in Lin et al. \cite{lin2021explicitly}, while large learning rates may cause instability or even NaN issues ruining the whole process.
In our numerous trials, we found that an initial learning rate of 0.01 with a slightly larger decay rate (5000 as opposed to the default 10000) hits a nice balance. 
It took millions of training steps to finally obtain a converge energy in FermiNet-DMC.
In comparison, FermiNet-VMC has not converged even after four million steps.
%
Our calculated energy for a T-shaped benzene dimer with bond length 4.95 \AA~is listed in Supplementary Table \ref{tab:benzene_dimer_energy}.
The DMC energy of dimers are still higher than the doubled monomer DMC energy by 2 mHa, indicating that a 3-layer network is too restricted to handle a benzene dimer, which would lead to a bias when estimating binding energies.
%

\begin{table}[htb]
\caption{The ground state energy in Hartree for a T-shaped benzene dimer with bond length 4.95 \AA.  VMC and DMC results are based on a FermiNet trained for $4\times 10^6$ steps.}\label{tab:benzene_dimer_energy}
\centering
\begin{tabular}{ll}
\toprule
Methods &  Energy ($E_{\rm h}$)\\
\midrule
FermiNet-VMC & -464.4067(3)\\
FermiNet-DMC & --464.4640(2)\\
\bottomrule
\end{tabular}
\end{table}

In principle, the binding energy can be accurately determined when the ground state energy of the dimer is calculated using a sufficiently large network, which however is beyond the capability of our available computational resource.
In order to reduce the error in estimating the binding energy, 
we train a separate FermiNet for a configuration with two benzene molecules separated at 10 \AA, which we dub as ``long-distance''.
The training curve is also displayed in Supplementary Fig. \ref{fig:bd_binding_energy_si}.a.
The calculated binding energy results, taken as the energy difference between the long-distance configuration and the equilibrium configuration, are shown in Fig. \ref{fig:bd_binding_energy_si}b.
Although FermiNet-DMC improves the convergence and the final results of FermiNet-VMC significantly, it still shows a slight overbinding compared with experimental measured range.
%
%
From the energy convergence pattern, the DMC energy for the long-distance configuration converges earlier than the one for the equilibrium configuration, resulting in an overly-large binding energy as the energy for equilibrium configuration gets smaller. 
%
The reason still lies in the inconsistency of two neural networks in describing two different configurations.
To begin with, a long-distance configuration may be more difficult for FermiNet to handle, especially when its expressive power is limited.
Similar phenomenon also shows up in \cite{luchow1996accurate} where DMC energy is too high for the long-distance configuration due to the limitation of the trial wavefunction.
Moreover, the neural networks for different configurations are trained separately and therefore they may converge to local minima of different quality introducing additional bias in the binding energy calculation. 

\begin{figure}[t]
  \centering
  \includegraphics[width=174mm]{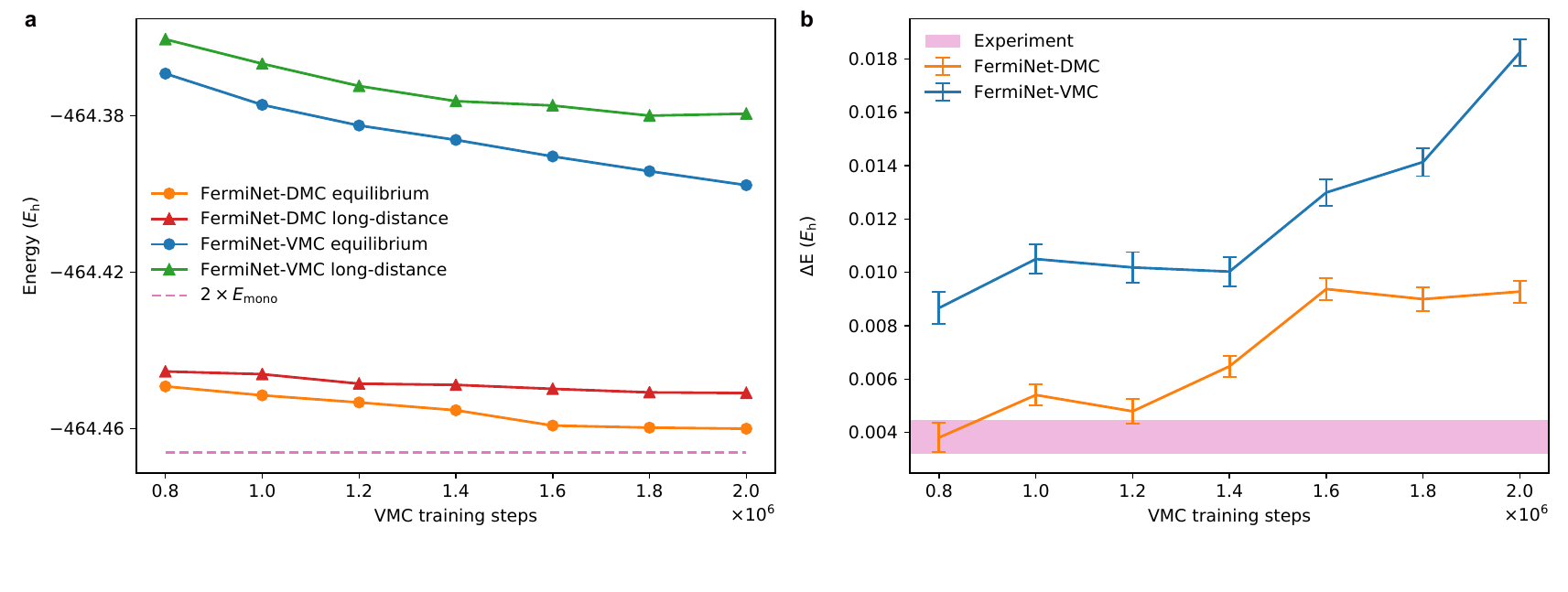}
  \caption{\label{fig:bd_binding_energy_si} \textbf{a}. The energy curve for both equilibrium configurations with bond length 4.95 \AA~and long-distance configurations with bond length 10 \AA~calculated with FermiNet-VMC and FermiNet-DMC. \textbf{b}. Benzene dimer binding energy calculated with both FermiNet-VMC and FermiNet-DMC. Both results exceed the experimental range while FermiNet-DMC result is closer to the experimental upper limit.}
\end{figure}

\section{Binding energy extrapolation}\label{sec:benzene-extrapolation}
In this note we provide details on the binding energy extrapolation for benzene dimer discussed in main text Section II G. 

We first show how to fit the slope parameter $w$ in main text Eq. 1.
For a given VMC training process, we do VMC inference and DMC process at a sequence of steps $\{k\}$ and collect VMC and DMC energy estimate $\{E_{\rm VMC}^{(k)}\}$ and $\{E_{\rm DMC}^{(k)}\}$. 
Following main text Eq. (1), we assume
\begin{equation}\label{eq:extrapolation_with_noise}
    E_{\rm DMC}^{(k)} - E_{\rm ex} = w\cdot (E_{\rm VMC}^{(k)} - E_{\rm DMC}^{(k)}) + b + \epsilon^{(k)}
\end{equation}
where $\epsilon^{(k)}$ corresponds to 0-centered random noise.

A simple manipulation of Supplementary Eq. \eqref{eq:extrapolation_with_noise} leads to 
 \begin{equation}\label{eq:bd_linear_fit}
    E_{\rm DMC}^{(i)} - E_{\rm DMC}^{(j)} =w \cdot [(E_{\rm VMC}^{(i)} - E_{\rm VMC}^{(j)} ) - (E_{\rm DMC}^{(i)} - E_{\rm DMC}^{(j)})] + (\epsilon^{(i)} - \epsilon^{(j)})  \quad \forall i, j
\end{equation}
which can be used to fit for slope $w$ with data $\{E_{\rm VMC}^{(k)}\}$ and $\{E_{\rm DMC}^{(k)}\}$ using least squares.
From main text Fig. 5c, we can tell that the slope of fitted line for benzene dimers with different bond length are very close. 
Therefore we choose to fit a single slope $w$ with the data from two configurations combined, as shown in Supplementary Fig. \ref{fig:bd_fit_w}. 

With the fitted slope $w_0$, we have extrapolation from Supplementary Eq. \eqref{eq:extrapolation_with_noise}
\begin{equation*}
    E_{\rm ex,~s} = (1 + w_0)\cdot E_{\rm DMC,~s}^{(i)} - w_0\cdot E_{\rm VMC,~s}^{(i)} - b - \epsilon^{(i)}_s\\
\end{equation*}
where $s$ denotes the bond length for the benzene dimer and can choose value from 4.95 and 10 in this case.
Therefore,
\begin{equation}\label{eq:delta_e_extrapolation_with_noise}
    \Delta E_{ex}=E_{\rm ex,~10} - E_{\rm ex,~4.95} = (1 + w_0) \cdot (E_{\rm DMC,~10}^{(j)} - E_{\rm DMC, ~4.95}^{(i)}) - w_0\cdot (E_{\rm VMC,~10}^{(j)} - E_{\rm VMC,~4.95}^{(i)}) - (\epsilon_{10}^{(j)} - \epsilon_{4.95}^{(i)})
\end{equation}
Namely, with this scheme, we can get the extrapolated binding energy $\Delta E_{ex}$ from energy difference from any pair of training steps, modulo the fitting error.
And this explains why the distribution of our extrapolated result concentrates so well in main text Fig. 5b.
By averaging the Supplementary Eq. \eqref{eq:delta_e_extrapolation_with_noise} over all chosen steps $i$ and $j$, we significantly reduce the fitting error and thus get a much more accurate binding energy estimate, also shown in main text Fig. 5b.

\begin{figure}[t]
  \centering
  \includegraphics[width=80mm]{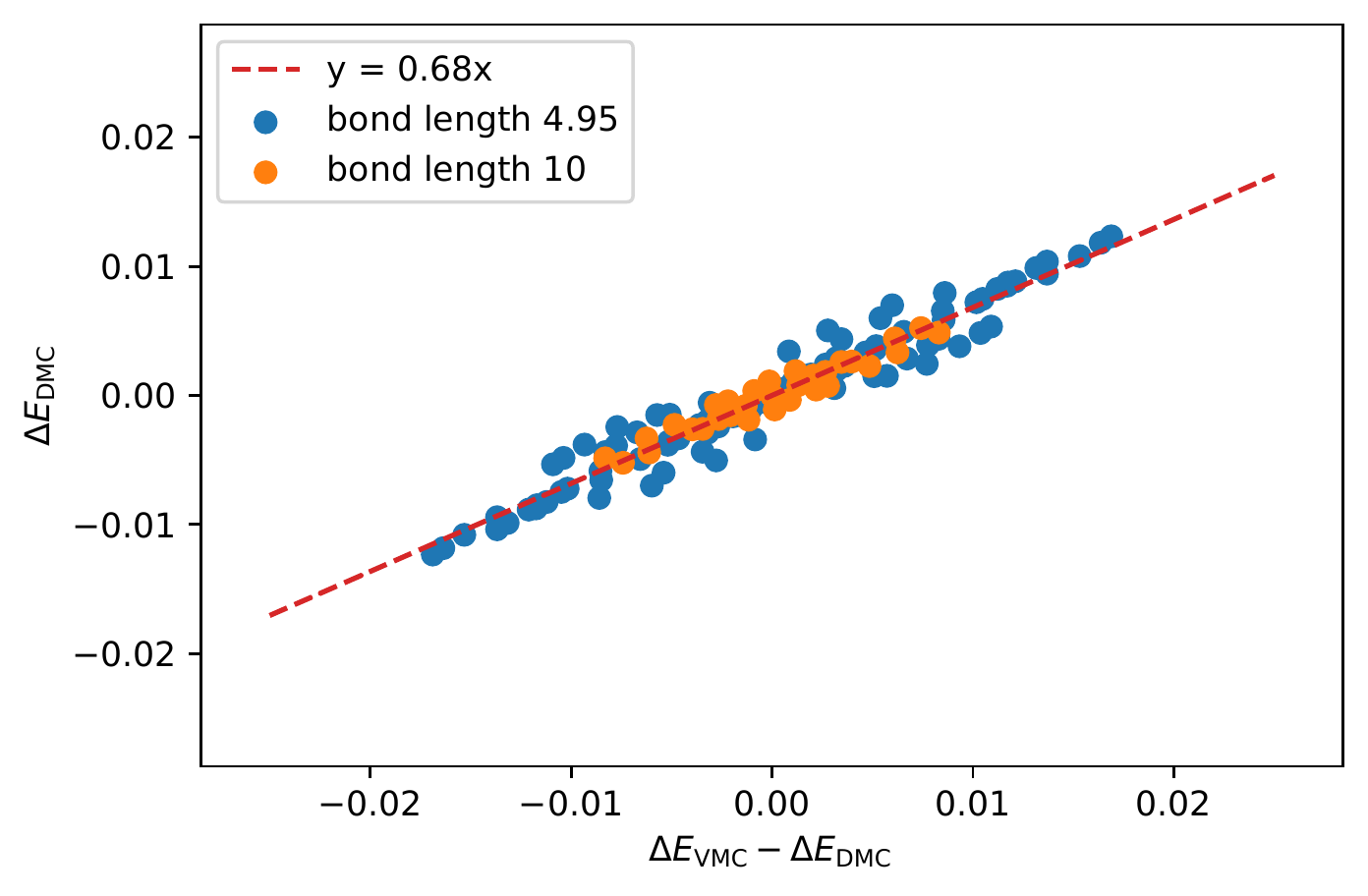}
  \caption{\label{fig:bd_fit_w}. For a pair of steps $i$ and $j$ in VMC training for benzene dimer, we plot two quantities showing up in Supplementary Eq. \eqref{eq:bd_linear_fit}, namely $[(E_{\rm VMC}^{(i)} - E_{\rm VMC}^{(j)} ) - (E_{\rm DMC}^{(i)} - E_{\rm DMC}^{(j)})]$ v.s. $E_{\rm DMC}^{(i)} - E_{\rm DMC}^{(j)}$. The data for benzene dimer with bond length 4.95 \AA~and 10 \AA~ have very close slope. With those two set of data combined, we fit a straight line with least squares.  
  }
\end{figure}

\section{CCSD(T) calculations and their extrapolations}
All CCSD(T) results in this paper are obtained with Psi4 \cite{psi4}. Four-center electron repulsions are approximated with density fitting. Instead of the usual Dunning basis sets (cc-pVXZ) for valence-only calculation, we used Dunning basis sets (cc-pCVXZ) for valence-core correlations which approximate the exact solution to the all-electron problem. Since the corresponded auxiliary basis set for cc-pCV5Z is not available in the current version of Psi4 (v1.6), we assigned cc-pV5Z-jkfit as the auxiliary basis for cc-pCV5Z basis. 

For the extrapolation of complete basis set (CBS) limit, we follow the exponential form \cite{halkier1999basis} \begin{equation}\label{eq:ccsdt_extrapolation}
    E(X) = E_\infty + B e^{-\alpha X}
\end{equation}
where $X$ is the cardinal number corresponding to the number of basis functions for each atomic orbital (e.g. 2 for cc-pCVDZ, 3 for cc-pCVTZ). $E_\infty$ is the energy at CBS limit. $B$ and $\alpha$ are parameters needed to fit. Hartree-Fock energy, CCSD correlation energy, and (T) contribution are fitted separately using the same formula. Then three extrapolated contributions are added to the CCSD(T)/CBS result. We denote CBS(2-4) and CBS(3-5) as the extrapolations using cc-pCVXZ (X=2,3,4), and cc-pCVXZ (X=3,4,5) respectively. The detailed calculation for benzene is summarized in Table \ref{tab:extrapolation}.
\begin{table}[htb]
\caption{Extrapolation of CBS limit for benzene molecule}\label{tab:extrapolation}
\centering
\begin{tabular}{llllll}
\toprule
basis &  HF & CCSD & CCSD(T) & $E_{corr}$ & (T) contribution\\
\midrule
cc-pCVDZ &	-230.724368&	-231.777896&	-231.815816&	-1.053528&	-0.037920 \\
cc-pCVTZ &	-230.779735&	-232.057476&	-232.112446&	-1.277741&	-0.054970 \\
cc-pCVQZ	& -230.793378 &	-232.143423&	-232.202980&	-1.350045&	-0.059557 \\
cc-pCV5Z &	-230.796283&	-232.167077&	-232.227835&	-1.370788&	-0.060758 \\
\midrule
CBS(2-4) &	-230.797839&		&-232.243544&	-1.384460&	-0.061245 \\
CBS(3-5) &	-230.797069&		&-232.237386&	-1.379133&	-0.061184 \\
\bottomrule
\end{tabular}
\end{table}

Our best estimate of CCSD(T)/CBS for benzene is $-232.237386$ Ha, given by CBS(3-5). The difference between CCSD(T)/CBS(2-4) and CCSD(T)/CBS(3-5) is about 6 mHa. We expect the extrapolation error of CBS(3-5) for Benzene molecule to be of the order of 1 mHa. The uncertainty of the extrapolation scheme is mostly contributed by the extrapolation of $E_{corr}$.  Hartree-Fock energy and (T) contribution converge fast with respect to the basis set. Their uncertainties (difference between CBS(2-4) and CBS(3-5)) are less than 1 mHa, and almost negligible. 

For Benzene dimer, the largest basis set we calculated is cc-pCVQZ. Hence, our best result for Benzene dimer is obtained by CBS(2-4), which gives $-464.486530$ Ha. We expect the extrapolation error of CBS(2-4) for Benzene dimer is of the order of $10$ mHa, due to the uncertainty of CBS(2-4) and the doubled system size compared to a single Benzene molecule. We also calculated CCSD(T)/CBS(2-4) binding energy $E_{\rm binding} = 2\times E_{\rm mono} - E_{\rm 4.95}$ where $E_{\rm mono}$ is the energy of benzene molecule, and $E_{\rm 4.95}$ is the energy of benzene dimer at 4.95 \AA. The binding energy is calculated for each finite basis set first. Then, they are extrapolated with Supplementary Eq. \eqref{eq:ccsdt_extrapolation}. The binding energy converges with respect to the basis set much faster than the absolute energy does (Table \ref{tab:binding_extrapolation}). The CBS(2-4) estimate for Benzene dimer is $2.66$ kcal/mol which is already consistent with the existing result \cite{miliordos2014benchmark,bd_exp}. 
\begin{table}[htb]
\caption{Extrapolation of CBS limit for binding energy of benzene dimer}\label{tab:binding_extrapolation}
\centering
\begin{tabular}{lll}
\toprule
basis & \multicolumn{2}{l}{CCSD(T) binding energy} \\
& (hartree) & (kcal/mol) \\
\midrule
cc-pCVDZ & 0.004855 &	3.046686 \\
cc-pCVTZ & 0.004659 &	2.923544 \\
cc-pCVQZ & 0.004526 &    2.839844 \\
\midrule
CBS(2-4) & 0.004243 &	2.662232 \\
\bottomrule
\end{tabular}
\end{table}
 Leveraging this fact, we are able to calculate the Benzene dimer CBS(3-5) result with Benzene monomer CBS(3-5) result and binding energy (2-4) result, which gives -464.479015 Ha. Namely $E_{\rm 4.95~(3-5)} = 2 \times E_{\rm mono~(3-5)} - E_{\rm binding~(2-4)}$. We believe this is more reliable than the CBS(2-4) result and report it in the main text Figure 5a. Here we don't consider the Basis Set Superposition Error (BSSE) because the difference between uncorrected and BSSE-corrected CBS estimates is negligible, according to \cite{miliordos2014benchmark}.

\section{Details on the divergence measuring nodal surface difference}
The neighborhood $S_\epsilon$ and mapping $\phi$ introduced in main text Section IV F for the wavefunction $\Psi$ and nodal surface S is difficult to compute in practice. Instead we propose easier-to-compute alternatives as approximation.

The major difficulty here is that given a point $y$ and a set $S$, it's hard to find the point $z\in S$ that is closest to $y$. 
We propose an approximation as follows. 
Note that the set $S$ of our interest is always a nodal surface associated with certain wavefunction $\Psi$. 
Instead of looking for $z\in S$ that is closest to $y$, we move $y$ in the direction in which the value of $\log(|\Psi|)$ decreases the fastest, namely we form the integral curve of the vector field  $-\nabla\log(|\Psi|)$ starting from $y$.
And we use the intersection of this integral curve and $S$ as the target point $z$, signaled by the change of sign of $\Psi$, and we approximate the distance from $y$ to $S$ as 
\begin{equation*}
    \tilde{d}(y, S) = d(y, z)
\end{equation*}
In other words, we are looking for $z\in S$ such that there exists curve $l(t)$ satisfying 
\begin{equation*}
    \begin{cases}  
    & \frac{dl}{dt} = -\nabla \log(|\Psi(l)|) \\
    & l(0) = y  \\
    & l(t_0) = z \in S
    \end{cases}  
\end{equation*}
where $t_0$ could be an arbitrary number, either positive or negative.
Denote the function mapping $y$ to $z\in S$ as $\xi$, and $\xi$ is used as our practical substitute for function $\phi$ mentioned in main text Section IV F looking for the closest point to y in $S$.
Function $\xi$ is well-defined in the neighbourhood of $S$, guaranteed by the local existence and uniqueness of ODE.


Similarly, we propose the alternative of $S_{\epsilon} = \{x | d(x, S) < \epsilon\}$ as 
\begin{equation*}
    \tilde{S}_\epsilon = \{x | \Psi(x) \cdot \Psi (x - \epsilon\cdot\frac{\nabla \log(|\Psi(x|)}{||\nabla \log(|\Psi(x)|)||}) < 0\}
\end{equation*}
where S is the nodal surface for wavefunction $\Psi$.
%

%
%


Our algorithm to compute the approximated divergence is in Supplementary Algo. \ref{algo:divergence}.
\begin{algorithm}[htp]
    \caption{Divergence for nodal surface difference pseudocode.}\label{algo:divergence}
    \KwData{wavefunction $\Psi$ for nodal surface S. wavefunction $\Phi$ for nodal surface T. A batch of walkers (W) following the distribution of normalized $\Psi^2$. The number of MCMC steps (M) to run in each iteration. $\epsilon$ controlling the size of neighborhood of nodal surfaces. $\eta$ controlling the discretization of integral curve when moving towards the nodal surface.}
    \KwOut{Divergence value D measuring difference between S and T.}
    \SetKw{KwCompute}{compute}
    Initialize set L as an empty set.
    \Comment{Sample points from $S_\epsilon$ according to $\Psi^2$}\\
    \While{length(L) < K} {
        \For{each walker $r_i$ in W}{  
              Run MCMC procedure from $r_i$ for M steps to get a new walker $q_i$.\\
              $u_i = q_i -  \epsilon\cdot\frac{\nabla \log(|\Psi(q_i)|)}{||\nabla \log(|\Psi(q_i)|)||}$\\
              \If{$\Psi(u_i) \cdot \Psi(q_i) < 0$} {
                Insert $q_i$ into L\\
              }
              $r_i \gets q_i$\\
        }
    }
    
    $ $\\
    Initialize set O as an empty set.
    \Comment{Move samples in L closer to node Surface $S$ of $\Psi$}\\
    \For{each walker in L }{  
              Move along the vector field $\frac{\nabla \log(|\Psi|)}{||\nabla \log(|\Psi|)||}$ with tiny step $\eta$ until the sign of $\Psi$ changes\\
              $o_i \gets$ the point right before the sign changes\\
              Insert $o_i$ into O\\
    }
    
    \Comment{Move samples in O closer to node surface $T$ of $\Phi$}\\
    \For{each walker $o_i$ in O }{  
              Move along the vector field $\frac{\nabla \log(|\Phi|)}{||\nabla \log(|\Phi|)||}$ with tiny step $\eta$ until the sign of $\Phi$ changes\\
              $p_i \gets $ the point right before the sign changes\\
              \KwCompute $d(o_i, p_i)$\\
    }
    \KwCompute D = Avg($d(o_i, p_i)$)
\end{algorithm}

\bibliography{ref}